\documentclass{elsart5}
\usepackage{graphicx}
\usepackage{amssymb}
\usepackage{bm}

\begin{document}

\begin{frontmatter}

\title{Spin Transfer Torques}

\author[aff1]{D. C. Ralph}
\ead{ralph@ccmr.cornell.edu}
\ead[url]{http://people.ccmr.cornell.edu/$\sim$ralph/}
\author[aff2]{M. D. Stiles}
\ead{mark.stiles@nist.gov}
\ead[url]{http://cnst.nist.gov/epg/epg\_home.html}
\address[aff1]{Laboratory of Atomic and Solid State Physics,
              Cornell University, Ithaca, New York 14853}
\address[aff2]{Center for Nanoscale Science and Technology, National
               Institute of Standards and Technology, Gaithersburg, Maryland
20899-6202}
\received{}
\revised{}
\accepted{}


\begin{abstract}
This tutorial article introduces the physics of spin transfer torques
in magnetic devices.  We provide an elementary discussion of the mechanism
of spin transfer torque, and review the theoretical and experimental progress
in this field.  Our intention is to be accessible to beginning
graduate students. This is the introductory paper for a cluster of
``Current
Perspectives'' articles on spin transfer torques published in
volume {\bf 320} of the
{\it Journal of Magnetism and Magnetic Materials}.  This article is meant to
set the stage for the others which follow it in this cluster;
   they focus in more depth on particularly interesting aspects of
spin-torque physics and highlight unanswered questions that might be
productive topics for future research.
\end{abstract}

\end{frontmatter}

\section{Introduction}\label{sec:intro}

The electrons that carry charge current in electronic circuits also
have spins.  In non-ferromagnetic samples, the spins are usually
randomly oriented and do not play a role in the behavior of the
device.  However, when ferromagnetic components are incorporated into
a device, the flowing electrons can become partially spin polarized
and these spins can play an important role in device function. Due to
spin-based interactions between the ferromagnets and electrons, the
orientations of the magnetization for ferromagnetic elements can
determine the amount of current flow.  By means of these same
interactions, the electron spins can also influence the orientations
of the magnetizations. This last effect, the so-called spin transfer
torque, is the topic of the following series of articles.  The goal
of these articles is to provide an introduction to the topical
scientific issues concerning the theories, experiments, and commercial
applications related to spin transfer torques.  These articles are written
by some of the leaders in the study of spin transfer torques and we
hope that they will be useful to beginning students starting their
study as well as of interest to experts in the field.

The authors of the succeeding articles were asked to
consider the field from their particular point of view, and to provide
a ``preview'' -- including discussion of interesting unsolved problems
-- rather than merely a review of completed work.  Hopefully, when
taken together these articles provide a broad and representative view
of where the field is and where it may be going.

The goal of the present article is to provide basic background and
context for the succeeding articles.
We start
with a very brief history of the field and provide references to
background material.  Then, in Section \ref{sec:ferro}, we introduce
aspects of ferromagnetism that are important to the discussion,
particularly for transition-metal ferromagnets, and discuss how spin
polarized currents arise.  Section \ref{sec:stt} describes how
spin transfer torques can be understood as resulting from changes in
spin currents.  Section \ref{sec:multi} explains some of the issues
associated with how spin transfer torques affect magnetic-multilayer
devices and tunnel junctions.  Section \ref{sec:wires} describes how
they affect domain walls in magnetic nanowires.  Finally, Section
\ref{sec:outlook} introduces the succeeding articles in the context
provided by the rest of this article.

The first work to consider the existence of spin transfer torques
occurred in the late 1970's and 1980's, with Berger's prediction that
spin transfer torques should be able to move magnetic domain walls
\cite{Berger:1978}, followed by his group's experimental observations
of domain-wall motion in thin ferromagnetic films under the influence
of large current pulses \cite{Freitas:1985,Hung:1988}.  The phenomenon
did not attract a great deal of attention at the time, largely because
very large currents (up to 45 A!) were required, given that the
samples were quite wide -- on the scale of mm.  However, with advances
in nanofabrication techniques, magnetic wires with 100 nm widths can
now be made readily, and these exhibit domain wall motion at currents
of a few mA and below.  Research on spin-transfer-induced domain wall
motion has been pursued vigorously by many groups since the early days
of the 21st century
\cite{Gan:2000,Ono:2001,Grollier:2002,Klaui:2003,Tsoi:2003}.
The present state of studies of spin-torque-driven domain
wall motion is described in the articles by Beach, Tsoi, and Erskine,
by Tserkovnyak, Brataas, and Bauer, and by Ohno and Dietl.

\begin{figure}
              \centering
              \resizebox{0.6\columnwidth}{!}{%
              \includegraphics{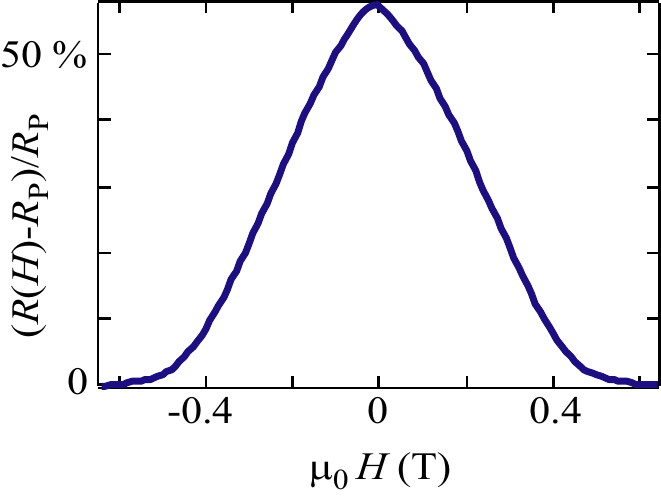}}
\caption{Giant Magnetoresistance (GMR) of a multilayer film with the
       structure (1.2~nm Co / 0.96~nm Cu)$_{10}$.  The measurement is done
       with the current flowing in the sample plane at a temperature of
       4.2~K.  The resistance is large near zero magnetic field where the
       interlayer exchange coupling between adjacent Co layers aligns
       their magnetizations antiparallel.  The resistance decreases as
       an applied magnetic field rotates the magnetizations to become
       parallel to each other.  $R_{\rm P}$ is the resistance when the two
       magnetizations are parallel.  Data are courtesy of Jordan Katine.
       }
              \label{fig:gmr}
\end{figure}

Widespread interest in magnetic nanostructures began in 1986 with the
discovery of interlayer exchange coupling
\cite{Grunberg:1986,Majkrzak:1986,Salamon:1986}.  Interlayer exchange
coupling \cite{iec_reviews} is the interaction between the
magnetizations of two ferromagnetic layer separated by an ultrathin,
non-ferromagnetic spacer layer.  Of particular interest was the discovery
of antiferromagnetic coupling in the Fe/Cr/Fe system by Gr\"unberg et
al. because it led shortly thereafter to the discovery of the Giant
Magnetoresistance (GMR) effect by Gr\"unberg's group and Fert's group
\cite{Baibich:1988,Binasch:1989}.  Gr\"unberg and Fert shared the 2007
Nobel Prize in Physics for this discovery.  GMR is the change in resistance
that occurs when the relative orientation of the magnetizations in two
ferromagnetic layers changes.  For example, when the magnetizations of
two Fe layers separated by Cr are antiparallel to each other as they
are when antiferromagnetically coupled, the sample has a relatively
large resistance.  The magnetizations of the Fe layers can be brought
into parallel alignment by an external magnetic field, and this
decreases the resistance.  At intermediate angles, the resistance is
also intermediate between these maximum and minimum values.  See
Fig.~\ref{fig:gmr}. The
overall size of the change in resistance is typically up to a few 10's
of percent, which is ``giant'' compared to the $\approx$ 1~\%
magnetoresistance changes of pure magnetic metals by themselves (due
to anisotropic magnetoresistance \cite{McGuire:1975}).  Most of the
early work on GMR focused on the sample geometry in which the current
flows in the plane (CIP) of the multilayer sample.  Another geometry,
first explored in 1991 \cite{Pratt:1991} has the current flow
perpendicular to the planes (CPP) of the multilayer and gives larger
fractional resistance changes \cite{cpp_reviews}.  The CPP geometry is
of particular interest in the present context because spin transfer
effects are more important than they are in the CIP geometry.

In studying the GMR effect, Parkin {\it et al}.\ discovered in 1990 that the
interlayer exchange coupling oscillates as a function of the thickness of the
spacer layer \cite{Parkin:1990}.  The oscillations could be quite
dramatic; up to sixty changes in sign of the coupling were seen in
single wedge-shaped samples allowing the simultaneous study of a range
of thicknesses \cite{Unguris:1994}.  Comparison of the oscillations
with calculations \cite{Bruno:1991} confirmed that this coupling was
an exchange interaction mediated by the electrons in the spacer layer
and that the oscillation periods were determined by the geometry of
the spacer layer's Fermi surface.

In 1989, Slonczewski \cite{Slonczewski:1989} calculated the interlayer
exchange coupling for the case in which the spacer layer is an
insulating tunnel barrier.  While there were no measurements of exchange
coupling across insulators at the time, his article has two features of
particular interest in regard to spin-torque physics.  First,
Slonczewski calculated the exchange coupling by determining the spin
current flowing through the tunnel barrier.
A spin current flows even with zero applied bias across the
tunnel junction whenever the magnetizations of the two electrodes are
non-collinear, and the source of the exchange coupling can be
understood to be the transfer of angular momentum from
this spin current to each magnet.
The second feature of
interest is that he considered the
additional coupling that results when a voltage is applied across the
junction.  This was the first calculation of a spin transfer torque in
a multilayer geometry with current flowing perpendicular to the plane.
However, there was little immediate experimental follow-up, because
the technology for making magnetic tunnel junctions at that time was
still rather primitive, and provided only tunnel barriers that were
too thick to permit the large current densities needed to excite
spin-torque-driven magnetic dynamics.

The papers most influential in launching the study of spin transfer
torques came in 1996, when Slonczewski \cite{Slonczewski:1996} and
Berger \cite{Berger:1996} independently predicted that current flowing
perpendicular to the plane in a metallic multilayer can generate a
spin transfer torque strong enough to reorient the magnetization in
one of the layers.  Since the metallic magnetic multilayers used for
GMR studies have low resistances (compared to tunnel barriers), they
could easily sustain the current densities required for spin transfer
torques to be important. Slonczewski predicted that the spin transfer
torque from a direct current could excite two qualitatively different
types of magnetic behaviors depending on the device design and the
magnitude of an applied magnetic field: either simple switching from
one static magnetic orientation to another or a dynamical state in
which the magnetization undergoes steady-state precession.  His
subsequent 1997 patent \cite{Slonczewski:patent} was remarkably
far-seeing, providing detailed predictions for many of the
applications that are currently being pursued.

\begin{figure}
              \centering
              \resizebox{0.9\columnwidth}{!}{%
              \includegraphics{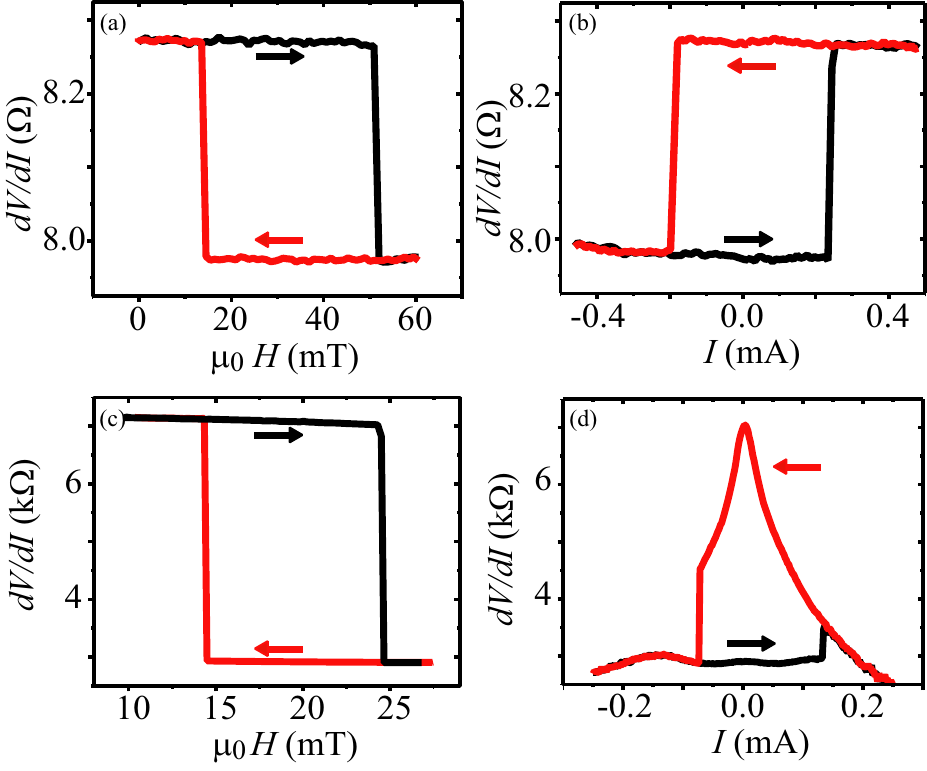}}
\caption{Comparison of magnetic switching at room temperature as
      driven by applied magnetic fields and by spin transfer torques.  (a)
      Switching for an all-metal nanopillar sample consisting of the
      layers 20 nm Ni$_{81}$Fe$_{19}$ / 12 nm Cu / 4.5 nm
      Ni$_{81}$Fe$_{19}$, as the magnetization of the thinner (free)
      magnetic layer is aligned parallel and antiparallel to the thicker
      magnetic layer by an applied magnetic field.  (b) Spin-torque-driven
      switching  by an applied current in the same device, with a constant
      magnetic field applied to give zero total field acting on the free
      layer.  (c) Switching for a magnetic tunnel junction nanopillar
      sample consisting of the layers 15~nm PtMn / 2.5~nm
      Co$_{70}$Fe$_{30}$ / 0.85~nm Ru / 3~nm Co$_{60}$Fe$_{20}$B$_{20}$ /
      1.25~nm MgO / 2.5~nm Co$_{60}$Fe$_{20}$B$_{20}$, as the 2.5-nm
      Co$_{60}$Fe$_{20}$B$_{20}$ free layer is reversed by an applied
      magnetic field.  (d) Spin-torque-driven switching  by an applied
      current in the same tunnel junction, with a constant magnetic field
      applied to give zero total field acting on the free layer.
Data for (a) and (b) are from \cite{Braganca:2005}, and data for (c)
      and (d) are courtesy of Jonathan Sun.
       }
              \label{fig:switching}
\end{figure}

\begin{figure}
              \centering
              \resizebox{0.9\columnwidth}{!}{%
              \includegraphics{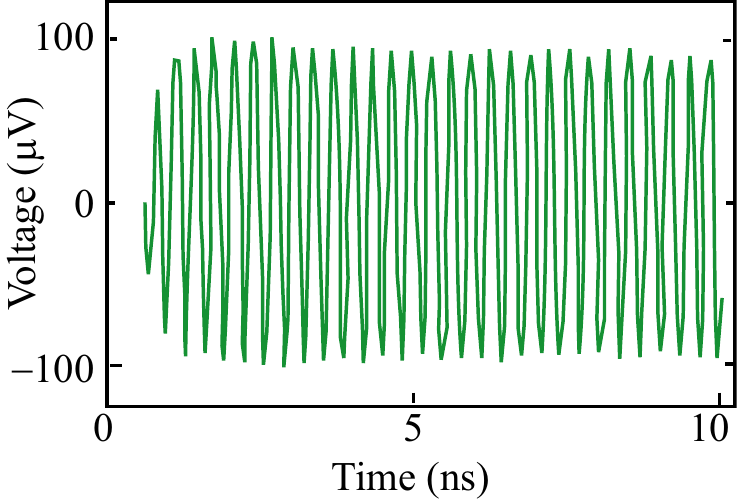}}
\caption{Voltage oscillations produced by steady-state precession of
       the magnetic free layer in a nanopillar sample, in response to the
       spin transfer torque from a 8.4~mA current step.  The sample had
       the layer structure 8~nm Ir$_{20}$Mn$_{80}$ / 4~nm
       Ni$_{80}$Fe$_{20}$ / 8~nm Cu / 4~nm Ni$_{80}$Fe$_{20}$, with the
       direction of the exchange bias from the Ir$_{20}$Mn$_{80}$ layer
       oriented 45$^{\circ}$ from the easy axis of the upper
       Ni$_{80}$Fe$_{20}$ free layer.  The measurement was made at a
       temperature of 40 K using a sampling oscilloscope
       \cite{Krivorotov:2005}.
       }
              \label{fig:oscil}
\end{figure}

Measurements of current-induced resistance changes in magnetic
multilayer devices were first identified with spin-torque-driven
excitations in 1998 by Tsoi {\it et al}., for devices consisting of a
mechanical point contact to a metallic multilayer \cite{Tsoi:1998},
and in 1999 by Sun in manganite devices \cite{Sun:1999}.  Observation
of magnetization reversal caused by spin torques in lithographically
defined samples occurred shortly thereafter
\cite{Myers:1999,Katine:2000}.  Fig.~\ref{fig:switching} shows
comparisons between spin-torque-driven magnetic switching and
magnetic-field driven switching for a metallic multilayer
and a magnetic tunnel junction.
Phase-locking between
spin-torque-driven magnetic precession and an alternating magnetic
field was detected in 2000 \cite{Tsoi:2000} and direct measurements of
steady-state high-frequency magnetic precession caused by spin torque
from a direct current were made beginning in 2003
\cite{Kiselev:2003,Rippard:2004,Krivorotov:2005}.
Figure~\ref{fig:oscil} shows an example of voltage
oscillations due to spin-torque-driven magnetic precession.
The article by
Berkov and Miltat describes some of these results and shows to what
extent it is currently possible to make a detailed comparison between
theory and experiment.  The article by Silva and Rippard discusses
some of the open questions pertaining to spin-torque-driven precession
in the point-contact sample geometry.

At the same time that interest was beginning to grow regarding spin
transfer torques in metallic multilayers, so was the interest in
magnetic tunnel junctions, starting with the observation in 1995 of
substantial tunnel magnetoresistance (TMR, the difference in
resistance between parallel and antiparallel orientation for the
electrode magnetizations of a magnetic tunnel junction) at room
temperature \cite{Miyazaki:1995,Moodera:1995}.  Most of the early
studies used aluminum oxide as a tunnel barrier.  Since aluminum oxide
barriers are heavily disordered it is difficult to study them
theoretically.  In 2001, Butler {\it et al}.\ and Mathon and Umerski
calculated the tunneling properties of the Fe/MgO/Fe system
\cite{Butler:2001,Mathon:2001}, which can be lattice-matched and
potentially well-ordered.  They found that the symmetry of the system
and the relevant electronic states lead to the possibility of an
extremely large TMR.  In 2004, values of TMR greater than those
observed for aluminum oxide barriers were published
\cite{Parkin:2004,Yuasa:2004} and the values of TMR demonstrated
experimentally have continued to increase rapidly until this day.

Techniques have now been developed to make both aluminum oxide and MgO
tunnel barriers sufficiently thin to support the current densities
needed to produce magnetic switching with spin transfer torques
\cite{Huai:2004,Fuchs:2004,Hayakawa:2005}. Work is underway to investigate
spin-torque-driven precession in tunnel barriers, as well.  One reason
for the interest in spin-torque effects in tunnel junctions is that
tunnel junctions are better-suited
than metallic magnetic multilayers for many types of
applications. Tunnel junctions have higher resistances that can often
be better impedance-matched to silicon-based electronics, and TMR
values can now be made larger than the GMR values in metallic
devices. The science and technology of spin transfer torques in tunnel
junctions are discussed in the articles by Katine and Fullerton and by
Sun and Ralph.

The possibility of commercial application has been a strong driving
force in this field from the beginning.  Gr\"unberg
\cite{Grunberg:patent} filed a German patent for applications
of GMR in 1988 even before the effect was published in the scientific
literature, and the phrase ``giant magnetoresistance'' now appears in
over 1500 US patents.  Devices based on the GMR and TMR effects have
already found very widespread application as the magnetic-field
sensors in the read heads of magnetic hard disk drives, and a
non-volatile random access memory based on magnetic tunnel junctions
has recently been introduced.  Slonczewski's 1997 patent
\cite{Slonczewski:patent} for devices based on spin transfer torques
has been referenced by 64 subsequent patents.  Applications of spin
transfer torques are envisioned using both types of magnetic dynamics
that spin torques can excite.  Magnetic switching driven by the
spin transfer effect can be much more efficient than switching driven
by current-induced magnetic fields (the mechanism used in the existing
magnetic random access memory).  This may enable the production of
magnetic memory devices with much lower switching currents and hence
greater energy efficiency and also greater device density than
field-switched devices.  The steady-state magnetic precession mode
that can be excited by spin transfer is under investigation for a
number of high-frequency applications, for example nanometer-scale
frequency-tunable microwave sources, detectors, mixers, and phase
shifters.  One potential area of use is for short-range chip-to-chip
or even within-chip communications. Spin-torque-driven domain wall
motion is also
under investigation for memory applications.  Parkin
has proposed a ``Racetrack Memory'' \cite{Parkin:patent} which
envisions storing bits of information using many domains arranged
sequentially in a magnetic nanowire and retrieving the information by
using spin transfer torques to move the domains through a read-out
sensor.  The article by Katine and
Fullerton discusses in detail the opportunities and challenges for
potential applications.

There are several books that can provide more background information
for this set of articles on spin transfer torques.  One resource is
the four volume series {\it Ultrathin Magnetic Structures} edited by
Heinrich and Bland \cite{ums}.  These volumes contain articles on
almost all aspects of magnetic thin films and devices made out of
them.  Another useful book, written at a more pedagogical level, is {\it
Nanomagnetism: Ultrathin Films, Multilayers and Nanostructures},
edited by Mills and Bland \cite{nanomagnetism}.  The three volume
series {\it Spin Dynamics in Confined Magnetic Structures} edited by
Hillebrands, Ounadjela, and Thiaville covers dynamical aspects of
magnetic nanostructures \cite{spindynamics}.  The five volume set,
Handbook of Magnetism and Advanced Magnetic Materials, edited by
Kronm\"uller and Parkin, covers the entire field of magnetism
including the topics of interest here \cite{handbook}.  {\it Concepts
in Spin Electronics}, edited by Maekawa, provides another recent
overview \cite{MaekawaConcepts}.  Finally, The Journal of Magnetism
and Magnetic Materials published a collection of review articles
including several on topics related to magnetic multilayers in Volume
200 \cite{beyond2000}.  Specific chapters in these books and other
review articles on spin transfer torques will be mentioned
throughout this article.

\section{The Basics of Ferromagnetism}\label{sec:ferro}

{\it The Origin of Ferromagnetism.} Ferromagnetism occurs when an
electron system becomes spontaneously spin polarized.  In transition
metals, ferromagnetism results from a balance between atomic-like
exchange interactions, which tend to align spins, and inter-atomic
hybridization, which tends to reduce spin polarization.  An accurate
accounting of both effects is quite difficult
\cite{Fulde:1995,Held:1998,Yang:2001}.  However, a qualitative
understanding is straightforward.  In isolated atoms, Hund's rules
describe how to put electrons into nearly degenerate atomic levels
to minimize the energy.
Hund's first rule says to maximize the spin, that is, to put in as many
electrons with spins in one direction into a partially filled atomic
orbital before you start adding spins in the other direction.  The
energy gain that motivates Hund's rule is that Pauli exclusion keeps
electrons with the same spin further apart on average, thereby
lowering the Coulomb repulsion between them.  This energy is called
the atomic exchange energy.  In accordance with Hund's first rule,
essentially all isolated atoms with partially filled orbital levels
have non-zero spin moments.  Non-zero values of orbital angular
momentum can also contribute to the magnetic moment of isolated
atoms.  In solids, on the other hand, electron
states on neighboring atoms hybridize and form bands.  Band formation
acts to suppress the formation of magnetic moments in two ways.
First, hybridization breaks spherical symmetry for the environment of
each atom, which tends to quench any orbital component of the
magnetic moment.
Second, band formation also inhibits spin polarization.  If one
starts with a system of unpolarized electrons and imagines flipping
spins to create alignment, then there is a kinetic-energy cost
associated with moving electrons from lower-energy filled band states to
higher-energy unoccupied band states.
As a
result, most solids are not ferromagnetic.  There are, of course,
exceptions.  For
example in materials with tightly bound 4f-orbitals, the hybridization
is so weak that those levels do become spin polarized much as they do
in the atomic state.  The transition metal ferromagnets iron, cobalt,
nickel, and their alloys, having partially filled d-orbitals, are the
exceptions of particular interest in
these articles.

The transition metal ferromagnets have both strong exchange splitting
and strong hybridization. The exchange splitting can stabilize a
spin-polarized
ferromagnetic state, even in the presence of band formation, by
generating a self-consistent shift of the majority-electron-spin band
states to lower energy than the minority-electron-spin states, so as
to more than compensate for the kinetic-energy cost associated with
the formation of the polarization.

{\it Models of Ferromagnetic materials.}
The local spin density approximation (LSDA)
\cite{Kohn:1965,vonBarth:1972,Gunnarsson:1976,Jones:1989} accurately
describes much of the important physics in these systems.  It treats
the atomic-like exchange and correlation effects in mean field theory
and treats the hybridization exactly.  Without any fitting parameters
it accurately predicts \cite{Janak} many of the properties of
transition metal ferromagnets like the magnetic moment.  In this
approach, the electron density and spin density are the fundamental
degrees of freedom and the wave functions are formal constructs that
allow calculation of the density.  As such, there is no formal
justification for using the LSDA wave functions as the physical wave
functions.  However, the wave functions are a solution to an accurate
mean field theory (LSDA) and in practice they can serve as a good
approximation to the real wave functions in many cases.  Many
calculations of spin transfer torques are based on using the wave
functions found from the LSDA; see in particular the article from
Haney, Duine, N\'u\~nez, and MacDonald for examples.

\begin{figure*}
              \centering
              \includegraphics{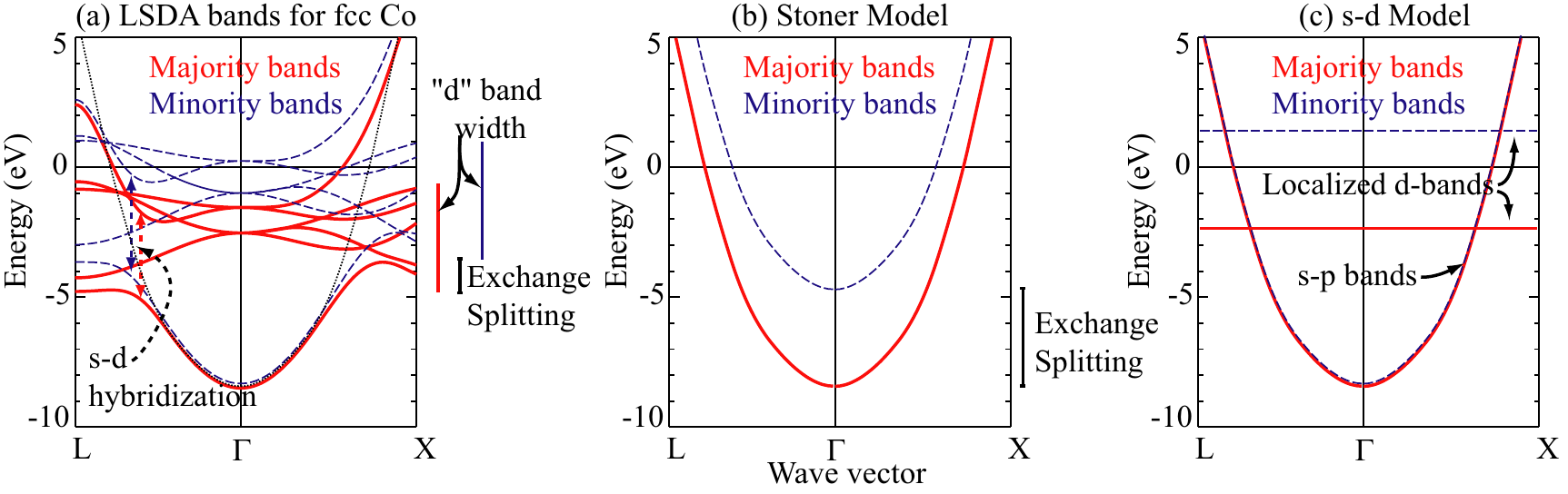}
\caption{Model band structures for ferromagnets.  The solid red (dashed blue)
      curves give the majority (minority) bands along
      two high symmetry directions through the Brillouin zone center,
      $\Gamma$.  Panel (a) gives bands calculated in the LSDA for
face-centered cubic (fcc) Co.
      The dotted black curve shows what the energy of the $s$-$p$ band
would be if it were
      not hybridized with the $d$ bands.  The bars to the right of (a)
      show the width of the $d$ bands and the shift between the
      majority and minority bands.  The dashed arrows in (a)
      indicated the widths of avoided level-crossings due to the hybridization
      between the $s$-$p$ and $d$ bands of the same symmetry along the
      chosen direction.
      Panel (b) gives a schematic version of a Stoner model for a
      ferromagnet.  The exchange splitting is larger than in (a) in order
to produce a
      reasonable size moment.  The majority and minority Fermi surfaces
      are more similar to each other than they are for the LSDA model.
      Panel (c) gives a schematic s-d model band structure.  The
      current-carrying s-p bands have a very small splitting due to
      the weak exchange interaction with the localized d-states.  The
      majority and minority Fermi surfaces are almost identical.
      }
              \label{fig:bs}
\end{figure*}

There are two simplified models of ferromagnetism that are sometimes
used to give
descriptions of the physics of spin transfer torques and for
calculations.  The first is the free-electron Stoner model.
This assumes that the electron
bands for spin-up and spin-down electrons have a
relative shift in energy due to an exchange interaction but otherwise
they both have a free-electron dispersion, $\epsilon({\bf k})= \hbar^2 k^2
/(2m) + \sigma_z \Delta/2$.  Here $\sigma_z$ is the Pauli spin matrix
and $\Delta$ is the exchange splitting.
The second simplified model is the s-d model. It was originally introduced
\cite{Langreth:1972} to describe local moment impurities in
a non-magnetic host.  The ``s'' electrons describe the delocalized
conduction band states of the host and the ``d'' electrons describe
the localized magnetic states, which are weakly coupled to the s
electrons.  Frequently, each d electron shell is treated as a local moment
${\bf S}$, which interacts with the conduction-electron spin density
${\bf s}$ through
a weak local interaction $-J{\bf S} \cdot {\bf s}$.  Neither the
Stoner Model nor the s-d model are well-justified approximations for
describing the transition-metal ferromagnets. See Fig.~\ref{fig:bs}
for a comparison of the band structures of these two models with a
more realistic band structure for Co
computed using the LSDA.  The simplified models can be very useful
for illustrating physical concepts, and sometimes for
estimates, but they are far from realistic.  The band structure in
transition-metal ferromagnets is considerably more complicated than
that of a single free-electron band as assumed in the Stoner model.
As for comparisons to the s-d model, in real materials the
hybridization within the d bands and of the d bands with the s bands are
quite strong, and so the d electrons cannot be considered localized.
On the other hand, the s-d model is one of several models that have
been used to
describe ferromagnetic semiconductors, like those discussed in the article
by Ohno and Dietl.  In these systems, the Mn substitutions are
believed to act very much like local moments.

The origins of spin-polarized currents in magnetic devices and the
giant magnetoresistance (GMR) effect can be understood as consequences
of the difference in band structures for the majority-spin and
minority-spin states in magnets, as predicted by LSDA calculations and
illustrated in a over-simplified way by the exchange splitting in the
free-electron Stoner model.  Spin-polarized currents can come about
because the spin-dependent electron properties in ferromagnets allow
magnetic thin films to act like spin filters.  Consider the example of
Cr/Fe multilayers in which GMR was first discovered.  For electrons
incident from Cr into a Fe layer, minority spins have a greater
probability to be transmitted through the Fe film than majority spins,
due to differences scattering caused by the
band-structure mismatch at the interface and also due to
band-structure-induced differences in the strength of scattering from
defects and impurities in the Fe layer \cite{Vouille:1999}.  Therefore
the current transmitted through the Fe layer in a Cr/Fe/Cr device is
partially spin-polarized in the minority direction, while the current
reflected from the layer is partially polarized in the majority
direction.  Many other factors can also affect the ultimate
polarization of the currents, including the layer thickness and the
spin flip scattering rates.

Once in a non-ferromagnetic metal like Cr or Cu, a spin-polarized current
persists on the scale of the spin diffusion length, typically on the
order of 100 nm or more in Cu and about 5 nm in Cr.  When two Fe
layers in a Cr/Fe
multilayer
are spaced more
closely than this and have parallel magnetic moments, minority
electrons have a high probability of being transmitted through both
layers, resulting in a relatively low overall device resistance even
though the majority electrons are scattered strongly.  In effect, the
minority electrons ``short out'' the structure, giving a low
resistance.  When nearby magnetic layers have antiparallel moments,
both majority and minority electrons scatter strongly in either one
layer or the other, and the resulting resistance is higher.  This is
the origin of GMR.  Other combinations of normal metals and magnetic
layers can act similarly, although for combinations like Cu/Co or
Cu/Ni it is the majority electrons that are more easily
transmitted through the magnetic layer, rather than the minority
electrons \cite{Vouille:1999,Bass:1999}.  A number of different
theoretical approaches have been used to calculate the transport
properties of magnetic multilayers and their GMR
\cite{Valet:1993,Brataas:2006,Stiles:2006}.  These techniques will be
discussed in more detail in Section \ref{sec:multi}.

{\it Micromagnetics.}
In order to describe the equilibrium configuration of the magnetization in a
ferromagnet, or the dynamical response to an applied magnetic field or a
spin-transfer torque,
it can be important to take into account that the magnetization
distribution may become spatially non-uniform.  Micromagnetics
\cite{Brown:1963,Fidler:2000,Aharoni:2001} is a phenomenological description of
magnetism on a mesoscopic length scale designed to model such
non-uniformities in
an efficient way.  It does not attempt to describe the behavior of the moment
associated with each atom,
but rather adopts a continuum description much like
elasticity theory.  Its utility arises because the length scales of
interest in magnetic studies are frequently much longer than atomic
lengths.  Atomic scale calculations become impractical in this case.

In equilibrium, the magnetization direction aligns
itself with an effective field, which can vary as a function of
position.  There are generally four main contributions to this effective field:
the externally applied magnetic field, magnetocrystalline
anisotropy, micromagnetic exchange, and the magnetostatic field.  Each of
the fields is most easily described in terms of an associated
contribution to the free energy.  The total effective field is then the
functional derivative of the free energy with respect to the magnetization
$\mu_0 H_{\rm eff}({\bf r})= - \delta E / \delta {\bf M}({\bf r}) $.
The magnetocrystalline anisotropy arises from the spin-orbit
interactions and tends to align the magnetization with particular
lattice directions.  Generally speaking the anisotropy field is a
local function of the magnetization direction and has a different
functional form for different lattices and materials.  The
micromagnetic exchange \cite{exchange} is the interaction that tends
to keep the magnetization aligned in a common direction, adding an
energy cost when the magnetization rotates as a function of position.
The magnetostatic interaction is a highly non-local interaction
between the magnetization at different points mediated by the magnetic
field produced by the magnetization.  Together, the four free energies can be
written
\begin{eqnarray}
\label{eq:mumage}
               E &=& -\mu_0 \int d^3 r {\bf H}_{\rm ext}\cdot{\bf M}({\bf r})
                    - { K_{\rm u} \over M_{\rm s}^2 }  \int d^3 r (\hat{\bf
n} \cdot{\bf
               M}({\bf r}))^2
            \\
                 && + { A_{\rm ex} \over M_{\rm s}^2 }
                  \int d^3 r \sum_\alpha ( {\partial\over \partial r_\alpha}
{\bf M} )^2
               \nonumber\\
                 && - { \mu_0 \over 8 \pi } \int d^3 r \int d^3 r'
                  {\bf M}({\bf r}) \cdot
                  { 3 ( {\bf M}({\bf r'}) \cdot {\bf x} ) {\bf x}
               - {\bf M}({\bf r'})|{\bf x}|^2 \over
                    |{\bf x}|^5 } , \nonumber
\end{eqnarray}
where ${\bf x}= {\bf r}-{\bf r}'$, $r_\alpha= x, y, z$, $M_{\rm s}$
is the saturation magnetization, $A_{\rm ex}$ is the exchange
constant, and $K_{\rm u}$ is the anisotropy constant.  Here we have taken
the specific example of a uniaxial anisotropy with an easy axis along
$\hat{\bf n}$.  The total effective field derived from
Eq.~(\ref{eq:mumage}) is
\begin{eqnarray}
\label{eq:mumagh}
               {\bf H}_{\rm eff} &=& {\bf H}_{\rm ext}
                 + { 2 K_{\rm u} \over \mu_0 M_{\rm s}^2 } \hat{\bf n}
(\hat{\bf n}
                 \cdot{\bf M}({\bf r}))
                 + { 2 A_{\rm ex} \over \mu_0 M_{\rm s}^2 }
                  \nabla^2 {\bf M}
               \nonumber\\
                 && + { 1 \over 4 \pi } \int d^3 r'
                  { 3 ( {\bf M}({\bf r'}) \cdot {\bf x} ) {\bf x}
               - {\bf M}({\bf r'})|{\bf x}|^2 \over
                    |{\bf x}|^5 } .
\end{eqnarray}
The atomic-like exchange, which drives the formation of the
magnetization and which is not explicit in these expressions,
places a strong energetic penalty on deviations of the magnitude of
${\bf M}({\bf r})$ away from $M_{\rm s}$.  This interaction is
generally taken into account by treating ${\bf M}({\bf r})$ as having
the fixed length $M_{\rm s}$.

{\it Magnetic Domains.}
The interactions within Eqs.~(\ref{eq:mumage}) and (\ref{eq:mumagh})
can compete with one another in determining the orientation of ${\bf
M}$ as a function of position. Different interactions can dominate on
different spatial scales, with the consequence that the magnetic
ground state is often spatially non-uniform, containing non-trivial
magnetization patterns even in equilibrium.  The micromagnetic
exchange and magnetocrystalline anisotropy both represent relatively
short-ranged or local interactions.  The micromagnetic exchange tends
to keep the magnetization spatially uniform and the magnetocrystalline
anisotropy can tend to keep it directed in particular lattice directions.
For the energy functional above, with uniaxial anisotropy, those
directions are $\pm\hat{\bf n}$.
For the materials of interest for spin transfer
applications the magnetocrystalline anisotropy is frequently weak and does not
play an
important role (although there are exceptions
\cite{Mangin:2006,Houssameddine:2007}).

Anisotropies in spin
transfer devices are
more commonly the result of the sample shape.  Magnetostatic
interactions
favor magnetization orientations aligned in the plane of thin-film samples and
along the long axis of samples with non-circular cross-sections.
Because of the dipole pattern of
long-ranged magnetic fields, the magnetostatic interaction can also favor
antiparallel alignment of the magnetization in distant parts of a
sample, and this can cause the magnetization pattern to become
non-uniform.  On short length scales the
magnetostatic interaction is relatively weak in comparison to micromagnetic
exchange.  However the
magnetostatic interaction is long-ranged so that it can eventually
dominate in large enough samples. This causes the magnetization
pattern to depend
on the sample size.
For magnetic thin-film
samples smaller than about 100~nm to 200~nm in diameter, the ground
state is approximately (but not exactly) uniform, with the
micromagnetic exchange dominating the magnetostatic interactions
\cite{Cowburn:2000}.  For samples
slightly bigger than this, magnetostatic interactions become more
important and the ground state can be a vortex state.
For still larger samples, the magnetization pattern may break up into
regions which have different magnetic orientations but within which
the magnetization is roughly uniform.  These regions are called
domains \cite{Hubert}.

The border region between domains is referred to as a domain wall.
Here the magnetization rotates over a relatively short distance from
one domain's orientation to the other's.  The more gradual is this
rotation, the less the cost in exchange energy.  However, wider domain
walls deviate from the low-energy orientation for magnetocrystalline
anisotropy over a larger volume, and therefore cost more anisotropy
energy.  The domain wall width is determined by a compromise
which minimizes the total energy of exchange + anisotropy, and can be
characterized roughly by the scale $\ell_{\rm DW}=\sqrt{A_{\rm
ex}/K_{\rm u}}$.  This length is strongly material dependent, ranging from
$\approx$~1~nm for hard magnetic materials to more than 100~nm for
soft magnetic materials.
In cases with weak anisotropy, domain wall widths are determined by
a competition between exchange and magnetostatic interactions.
The domain wall width can also depend on
sample geometry, and in a narrow contact between electrodes the wall
width can be narrowed in proportion to the contact diameter
\cite{Bruno:1999}.

In thin-film wires, typical of those used to study current-induced
domain wall motion, domain walls typically take one of two structures,
``transverse'' walls or ``vortex'' walls.  On either side of the wall,
the magnetization lies in-plane and points along the length of the
wire to minimize the magnetostatic energy, and there is a net
180$^{\circ}$ rotation of the magnetization at the wall.  In a
transverse wall, the magnetization simply rotates in the plane of the
sample from one domain to the other.  However, the competition between
the micromagnetic exchange energy and the magnetostatic energy causes
the wall width to be narrow (on the scale of the exchange length,
$\ell_{\rm ex}= \sqrt{2 A_{\rm ex}/(\mu_0 M_{\rm s}^2)} \approx $ 4~nm
to 8~nm, where $A_{\rm ex}$ is the prefactor of the micromagnetic
exchange in Eq.~(\ref{eq:mumage}) and $M_{\rm s}$ is the saturation
magnetization) on one edge of the wire and wide (on the scale of the
width of the wire) on the other edge.  A vortex wall is even more
complicated.  Here, the magnetization wraps around a central
singularity \cite{vcore}, the vortex core, giving a circulating pattern to the
magnetization.  The competition between these two wall
structures is
studied in Ref.~\cite{McMichael:1997}. Beach, Tsoi, and Erskine
describe how the detailed structure of domain walls plays a crucial
role in their motion when they are driven by either an applied
magnetic field or a spin transfer torque.

{\it Magnetic Dynamics in the Absence of Spin Transfer Torques.} When
a magnetic configuration is away from equilibrium, the magnetization
precesses around the instantaneous local effective field.
In the absence of dissipation, the
magnetization distribution stays on a constant energy surface.  In
order to account for energy loss, Landau and Lifshitz \cite{LL}
introduced a phenomenological damping term into the equation of motion
and Gilbert \cite{Gilbert} introduced a slightly different form
several decades later.  Both forms of the damping move the local
magnetization vector toward the local effective field direction:
\begin{eqnarray}
\label{eq:llg}
{\dot{\bf M}}
&=&               -\gamma_0' {\bf M} \times {\bf H}_{\rm eff}
                               - { \lambda \over M_s} {\bf M} \times
                                 \left( {\bf M} \times {\bf H}_{\rm 
eff}  \right)
\nonumber \\
&& {\mbox {\rm (Landau-Lifshitz)}}  \nonumber \\
{\dot{\bf M}} &=&               -\gamma_0 {\bf M} \times {\bf H}_{\rm eff}
                               + {\alpha \over M_s} {\bf M} \times 
{\dot{\bf M}},
\nonumber \\
&& {\mbox {\rm (Gilbert)}}
\end{eqnarray}
where $\gamma_0$ is the gyromagnetic ratio, $\lambda$ is the
Landau-Lifshitz damping parameter, and $\alpha$ is the Gilbert damping
parameter.  These two forms are known to be equivalent with the
substitutions, $\gamma_0' = \gamma_0 / (1 + \alpha^2)$ and $\lambda =
\gamma_0 \alpha / (1 + \alpha^2)$.  In spite of this equivalence,
there has been an ongoing debate about which is more correct.  This
debate has been rekindled with the interest in current-driven domain
wall motion (one of the present authors is guilty of contributing to
the debate) and is mentioned in the articles by
Tserkovnyak, Brataas, and Bauer and by Berkov and Miltat.  Part of
the fervor of the
debate arises from the fact that it is not experimentally testable.
Appropriate equations of motion can be formulated with either form of
damping at the expense of slight modifications to other terms in the
equation of motion.  This point is discussed further in
Sec.~\ref{sec:wires}.  Note that both the precession and damping terms
rotate the magnetization, but do not change its length.  This is
consistent with treating the magnetization as having a fixed length.

There have been many attempts to compute the damping parameters from
models for various physics processes \cite{Heinrich:2005}.  Some
mechanisms are intrinsic to the material, such as those due to
magnetoelastic scattering \cite{Rossi:2005}, and others are considered
extrinsic like two-magnon scattering from inhomogeneities
\cite{McMichael:2003}.  It appears that a model due to Kambersky
\cite{Kambersky:1976} for electron-hole pair generation describes the
dominant source of intrinsic damping in a variety of metallic systems
including
the ferromagnetic semiconductors \cite{Sinova:2004} and transition
metals \cite{Gilmore:2007} primarily of interest for spin transfer
torque applications.  For a magnetic element in metallic contact with
other materials, there can also be a contribution to the damping from
``spin-pumping'' -- the emission of spin-angular momentum from the
precessing magnet via the conduction electrons
\cite{Tserkovnyak:2002,Mizukami:2002,Heinrich:2003}.

Magnetization dynamics is most easily investigated using the macrospin
approximation.  The macrospin approximation assumes that the
magnetization of a sample stays spatially uniform throughout its
motion and can be treated as a single macroscopic spin.  Since the
spatial variation of the magnetization is frozen out, exploring the
dynamics of magnetic systems is much more tractable using the
macrospin approximation than it is using full micromagnetic
simulations.  The macrospin model makes it easy to explore the phase
space of different torque models, and it has been a very useful tool
for gaining a zeroth-order understanding of spin-torque physics.
However, for many systems of interest, even ones with very small
magnetic elements, the macrospin approximation breaks down.  For a
full understanding of magnetic dynamics, a micromagnetic approach is
therefore necessary.  The article by Berkov and Miltat discusses some
of the circumstances in which the macrospin approach is and is not a
reasonable approximation to the true dynamics.

For analyses of magnetic domain wall dynamics, there is a different
simplified description that can sometimes be useful for qualitative
understanding, as an alternative to full micromagnetic
calculations. The Walker ansatz \cite{Schryer:1974} restricts the
variation in the domain wall to uniform translations, uniform rotations
out of plane, and in some versions a uniform scaling of the wall width.
For the simplest version, the dynamics of the domain wall can then be
described by two degrees of freedom.  The article by
Beach, Tsoi and Erskine discusses when this approximation is valid and
when it is not for current-induced domain wall motion.

In some situations it is useful to consider the dynamics of a magnetic
sample in terms of the normal modes of the system, known as spin waves
or magnons, instead of directly integrating the equations of motion at
every point on a closely spaced grid designed to model the sample, as
is usually done in micromagnetic calculations. A spin wave is a small
amplitude oscillation of the magnetization around its average
direction. In macroscopic magnetic systems, the spectrum of spin waves
is essentially continuous as a function of frequency, but when
thin-film magnetic elements are shrunk to the scale of 100 nm in
diameter, the spin wave spectrum becomes measurably discrete
\cite{McMichael:2005}. In fact, the recent development of
spin-transfer-driven ferromagnetic resonance has made it possible to
measure the frequencies of these normal modes within individual
nanostructures \cite{Tulapurkar:2005,Sankey:2006}.  Interestingly,
calculations show that uniform precession, of the type assumed in the
macrospin approximation, is generally not a true normal mode because
the magnetostatic field is generally not uniform across the sample.
An analysis in terms of the normal modes can provide a strategy for
simulating magnetic dynamics that is more efficient than standard
micromagnetic simulations and somewhat more accurate than the
macrospin approximation.  The dynamics of a magnetic excitation can be
approximated by expanding the excitation using a finite set of normal
modes as a basis set, and determining the time dependence based on the
dynamics of the individual modes and their nonlinear couplings, rather
than by integrating the full equation of motion directly.  In cases
when the lowest-frequency most-spatially uniform normal mode
dominates the magnetic dynamics, the results are typically very
similar to the predictions of the simplest macrospin descriptions.
The relevance of the spin wave modes is discussed in the articles by
Berkov and Miltat and by Sun and Ralph.

\section{Spin Current, Spin Transfer Torque, and Magnetic
Dynamics}\label{sec:stt}

Thus far we have discussed the dynamics of magnets in the absence of
the spin transfer torque.  Spin transfer torques arise whenever the
flow of spin-angular momentum through a sample is not constant, but
has sources or sinks. This happens, for example, whenever a spin
current (created by spin filtering from one magnetic thin film) is
filtered again by another magnetic thin film whose moment is not
collinear with the first.  In the process of filtering, the second
magnet necessarily absorbs a portion of the spin angular momentum that
is carried by the electron spins. Changes in the flow of spin angular
momentum also occur when spin-polarized electrons pass through a
magnetic domain wall or any other spatially non-uniform magnetization
distribution.  In this process, the spins of the charge carriers
rotate to follow the local magnetization, so the spin vector of the
angular momentum flow changes as a function of position.  In either of
these cases, the magnetization of the ferromagnet changes the flow of
spin angular momentum by exerting a torque on the flowing spins to
reorient them, and therefore the flowing electrons must exert an equal
and opposite torque on the ferromagnet.  This torque that is applied
by non-equilibrium conduction electrons onto a ferromagnet is what we
will call the spin transfer torque.  Its strength can be calculated
either by considering directly the mutual precession of the electron
spin and magnetic moment during their interaction (an approach
discussed in the article by Haney,
Duine, N\'u\~nez, and MacDonald) or by considering the net change in
the spin current before and after the interaction (the approach we
will emphasize).

Our discussion in this Section will consist of two parts.  First we
will consider how it is that a spin-polarized current can apply a
torque to a ferromagnet.  This will be straightforward -- since a
torque is simply a time rate of change of angular momentum,
considerations of angular momentum conservation can be used to relate
the spin transfer torque directly to the angular momentum lost or
gained by spin currents.  We will use two simple toy models to
illustrate some of the physics involved in this process.  The second
part of our discussion will describe how to incorporate the spin
transfer torque into the equation of motion for the magnetization
dynamics.  This step of the argument will involve some more-subtle
points, related to the connection between the magnetization of a
ferromagnet and its total angular momentum.  To explore these points
fully, we will consider how one might derive the equation of motion
for the magnetization, $d{\bf M}/dt$, within a rigorous quantum
mechanical theory.

{\em Definition of the Spin Current Density.}
The primary quantity on which we will focus our interest will be the spin
current density ${\bf Q}$.  This has both a direction in spin space
and a direction of flow in real space, so it is a tensor
quantity.  For a single electron, the spin current is given
classically by the outer product of the average electron velocity and
spin density ${\bf Q}={\bf v}\otimes{\bf s}$.  For a single-electron
wavefunction $\psi$, the spin current density may be written
\begin{eqnarray}
             {\bf Q} &=& {\hbar^2 \over 2m} {\rm Im}
             (\psi^* {\bm\sigma} \otimes {\bm\nabla} \psi),
\end{eqnarray}
where $m$ is the electron mass, and
${\bm\sigma}$ represents the Pauli matrices $\sigma_x$, $\sigma_y$,
and $\sigma_z$.  The form of the spin current density is similar to
the more-familiar probability current density $(\hbar/m){\rm
Im}(\psi^*{\bf \nabla} \psi)$.  For a spinor plane-wave wavefunction
of the form
\begin{eqnarray}
\psi = {e^{ikx} \over \sqrt{{\Omega}}} \left( a\left|\uparrow \right\rangle + b
\left| \downarrow \right\rangle \right),
\end{eqnarray}
where ${\Omega}$ is a normalization volume, the spatial part of the spin
current points in the ${\bf
\hat{x}}$ direction, and the three spin components take the simple
forms
\begin{eqnarray}
\label{eq:qs}
Q_{xx} & = & {\hbar^2 k \over 2m{\Omega}} 2 {\rm Re}(ab^*) \nonumber \\
Q_{xy} & = & {\hbar^2 k \over 2m{\Omega}} 2 {\rm Im}(ab^*) \\
Q_{xz} & = & {\hbar^2 k \over 2m{\Omega}} (|a|^2 - |b|^2). \nonumber
\end{eqnarray}

By conservation of angular momentum, one can say that the
spin transfer torque acting on some volume of material can be
computed simply by determining the net flux of non-equilibrium spin
current through the surfaces of that volume, or equivalently by
integrating the divergence of the spin current density within an
imaginary pillbox surrounding the volume in question:
\begin{eqnarray}
\label{eq:pillbox}
               {\bf N}_{\rm st} & =& - \int_{{\rm pillbox}\ {\rm surfaces}} d^2
R \hat{\bf
n} \cdot {\bf Q} \nonumber \\
                          & =& - \int_{\rm pillbox\ volume} d^3 r \nabla
\cdot {\bf Q},
\end{eqnarray}
where ${\bf R}$ is the in-plane position and $\hat{\bf
n}$ is the interface normal for each surface of the pillbox. (Note
that since ${\bf Q}$ is a tensor, its dot
product with a vector in real space leaves a vector in spin space.)
If one prefers to think in terms of the differential form of
Eq.~(\ref{eq:pillbox}), it states that the spin torque density is the
divergence of the spin current density.

\begin{figure}
              \centering
              \resizebox{0.9\columnwidth}{!}{%
              \includegraphics{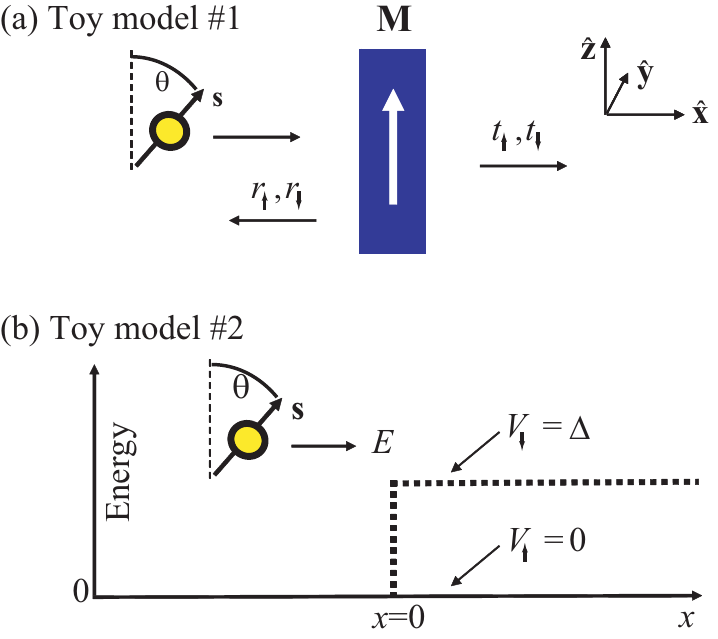}}
              \caption{Illustration of toy models discussed in the text.}
              \label{fig:toy}
\end{figure}

{\em Toy Model \#1.} Our first simple model is meant to illustrate that
when a spin polarized current interacts with a thin ferromagnetic layer
and undergoes spin filtering the result, in general, is that a
spin transfer torque is applied to the magnetic layer.
Consider the problem of a single-electron state with wave vector ${\bf
k}$ in the ${\bf \hat{x}}$ direction and spin oriented in the ${\bf
\hat{x}}$-${\bf \hat{z}}$ plane at an angle $\theta$ with respect to
the ${\bf \hat{z}}$ direction, which is incident onto a thin magnetic
layer whose magnetization is pointed in the ${\bf \hat{z}}$ direction
(see Fig.~\ref{fig:toy}(a)).  We will initially not be concerned about
what goes on
inside the magnetic layer, but we will account for its spin filtering
properties simply by assuming that it can be described by overall
transmission and reflection amplitudes for spin-up electrons
($t_{\uparrow}$, $r_{\uparrow}$) that are different from the
transmission and reflection amplitudes for spin-down electrons
($t_{\downarrow}$, $r_{\downarrow}$), and that no spin-flipping
processes occur.  Under these assumptions, the incident part of the
wavefunction is
\begin{eqnarray}
\label{eq:wavefunction}
\psi_{\rm in} = {e^{ikx} \over \sqrt{{\Omega}}}
\Big( \cos(\theta/2) \left|
\uparrow \right\rangle
+\sin(\theta/2) \left|\downarrow \right\rangle \Big).
\end{eqnarray}
This can be derived, for example, by starting with the
$\left|\uparrow\right\rangle$ state and applying the appropriate
rotation matrix for a spin-1/2 system \cite{Sakurai}.  The transmitted
and reflected parts of the scattering wavefunction are
\begin{eqnarray}
\psi_{\rm trans} &=& {e^{ikx} \over \sqrt{{\Omega}}} \Big(t_{\uparrow}
\cos(\theta/2) \left| \uparrow
\right\rangle + t_{\downarrow} \sin(\theta/2) \left|\downarrow
\right\rangle \Big) \nonumber \\
\psi_{\rm refl} &=& {e^{-ikx}\over \sqrt{{\Omega}}} \Big( r_{\uparrow}
\cos(\theta/2) \left| \uparrow
\right\rangle + r_{\downarrow} \sin(\theta/2) \left|\downarrow \right\rangle
\Big).
\end{eqnarray}
The components of the spin current density can be determined using
the expressions
given in Eq.~(\ref{eq:qs}).  The flows of spin density in the ${\bf
\hat{x}}$ spatial direction for the incident,
transmitted, and reflected parts of the wavefunction take the forms
\begin{eqnarray}
{\bf Q}_{\rm in} &=& {\hbar^2 k \over 2m{\Omega}} \Big[\sin(\theta) {\bf
\hat{x}}+ \cos(\theta) {\bf \hat{z}}\Big] \nonumber \\
{\bf Q}_{\rm trans} &=&
{\hbar^2 k \over 2m{\Omega}}
\sin(\theta){\rm Re}(t_{\uparrow}t_{\downarrow}^*) {\bf \hat{x}}
\nonumber\\
&&+
{\hbar^2 k \over 2m{\Omega}}
\sin(\theta)
{\rm Im}(t_{\uparrow}t_{\downarrow}^*) {\bf \hat{y}}
\nonumber\\
&&+
{\hbar^2 k \over 2m{\Omega}}
\Big[ |t_{\uparrow}|^2
\cos^2(\theta/2)- |t_{\downarrow}|^2 \sin^2(\theta/2)\Big] {\bf \hat{z}} \\
{\bf Q}_{\rm refl} &=&
-{ \hbar^2 k \over 2m{\Omega}}
\sin(\theta){\rm Re}(r_{\uparrow}r_{\downarrow}^*) {\bf \hat{x}}
\nonumber\\
&&
-{ \hbar^2 k \over 2m{\Omega}}
             \sin(\theta)
{\rm Im}(r_{\uparrow}r_{\downarrow}^*) {\bf \hat{y}}
\nonumber\\
&&
-{ \hbar^2 k \over 2m{\Omega}}
              \Big[ |r_{\uparrow}|^2
\cos^2(\theta/2)- |r_{\downarrow}|^2 \sin^2(\theta/2)\Big]
{\bf\hat{z}}. \nonumber
\end{eqnarray}
It is then clear that the total spin current is not conserved during
the filtering process: the spin current density flowing on the
left of the magnet ${\bf Q}_{\rm in} + {\bf Q}_{\rm refl}$ is not
equal to the spin current density on the right ${\bf Q}_{\rm trans} $.
By Eq.~(\ref{eq:pillbox}), we can say that the spin transfer torque
${\bf N}_{\rm
st}$ on an area $A$ of the ferromagnet is equal to the net spin current
transferred from the electron to the ferromagnet, and is given in
this toy model by
\begin{eqnarray}
                 {\bf N}_{\rm st}
                 &=& A {\bf \hat{x}} \cdot ({\bf Q}_{\rm in} + {\bf 
Q}_{\rm refl}
- {\bf Q}_{\rm
trans}) \nonumber\\
                   &=& {A \over \Omega}{\hbar^2 k \over 2m} \sin(\theta) \Big[1-
{\rm Re}(t_{\uparrow}t_{\downarrow}^*  + r_{\uparrow}r_{\downarrow}^*) \Big]
{\bf
\hat{x}}
\nonumber\\
&&- {A \over \Omega}{\hbar^2 k \over 2m}
\sin(\theta) {\rm Im}(t_{\uparrow}t_{\downarrow}^*  +
r_{\uparrow}r_{\downarrow}^*) {\bf
\hat{y}} .
\label{eq:Model1}
\end{eqnarray}
We have used the fact that $|t_{\uparrow}|^2 + |r_{\uparrow}|^2 = 1$ and
$|t_{\downarrow}|^2 + |r_{\downarrow}|^2 = 1$. There is no component
of spin torque in the ${\bf \hat{z}}$ direction. We find the general
result that the spin transfer torque is zero when $t_{\uparrow} =
t_{\downarrow}$ and $r_{\uparrow} = r_{\downarrow}$ (in which case the
``magnetic'' layer would provide no spin filtering) or when the
incoming spin orientation is collinear with the magnetization of the
layer, $\theta = 0$ or $\pi$.  However, for any non-collinear
spin orientation, when the magnet does provide spin filtering,
it is a direct consequence of the spin filtering that the spin
transfer torque acting on the ferromagnetic layer is non-zero.  This
torque is perpendicular to the magnetization of the layer (no ${\bf
\hat{z}}$ component), and for an individual incident electron the
torque may have components in both the ${\bf \hat{x}}$ and ${\bf
\hat{y}}$ directions, depending on the values of the transmission and
reflection coefficients.

It is possible to extend this type of 1-d toy model to calculate the
torque applied to a magnetic thin film in a realistic 3-dimensional
sample.  The calculation proceeds by summing the torque contributed
by electron waves incident
onto the magnetic thin film from throughout the Fermi surface of the
non-magnetic metal, corresponding to electrons incident from many
directions in real space \cite{Waintal:2000}.  This requires summing
the contributions to the ${\bf \hat{x}}$ and ${\bf \hat{y}}$
components of the torque in Eq.~(\ref{eq:Model1}).  In terms of the
terminology commonly used in this field \cite{Brataas:2000,Brataas:2006},
the sum over the ${\bf \hat{x}}$-component contributions is
proportional to the real part of the ``mixing conductance'' and gives
the ``in-plane'' torque (the plane defined by the moments on the
ferromagnet and the incoming spin), and the sum over the ${\bf
\hat{y}}$-component contributions is proportional to the imaginary
part of the mixing conductance and gives a perpendicular torque.

{\em Toy Model \#2.} Next we consider a second simple 1-dimensional toy model
\cite{Stiles:2002a}, to
illustrate some of the processes that occur near a normal
metal/ferromagnetic interface and influence the spin torque.  Here we again
first assume a single incoming spin-polarized electron wavefunction of the form
given
by Eq.~(\ref{eq:wavefunction}) incident onto a magnetic layer whose
magnetization is in the ${\bf \hat{z}}$ direction.  However, in this
case we will use a Stoner-model approach to describe the magnetic
layer.  That is, we will assume that the electrons inside the
ferromagnetic layer experience an exchange splitting $\Delta$ which
shifts the states in the minority-spin band (down electrons) higher in
energy than the majority-spin band (up electrons), but that both bands
have a free-electron dispersion.  The physics near the interface can
then be modeled as a simple scattering problem in which the electron
scatters from a rectangular potential-energy step (at position $x=0$)
that has different heights for spin-up and spin-down electrons
(see Fig.~\ref{fig:toy}(b)).  For
simplicity, we will assume that the height of the potential-energy
step is 0 for up spins and $\Delta$ for down spins, and we will
consider an electron energy $E = \hbar^2 k^2/(2m)$ which is greater
than $\Delta$.

By matching wavefunctions and their derivatives at the
interfaces, it is an elementary problem to calculate the transmitted
and reflected parts of the scattering-state wavefunction with energy
eigenvalue $E$:
\begin{eqnarray}
\psi_{\rm trans} &=& {e^{ik_{\uparrow} x} \over \sqrt{{\Omega}}}
\cos(\theta/2) \left|
\uparrow \right\rangle + {e^{ik_{\downarrow} x} \over \sqrt{{\Omega}}} { 2
k \over k + k_{\downarrow} }\sin(\theta/2)
\left|\downarrow \right\rangle  \nonumber \\
\psi_{\rm refl} &=& {e^{-ikx}\over \sqrt{{\Omega}}} {k -
k_{\downarrow} \over k +
k_{\downarrow} }
\sin(\theta/2) \left|\downarrow \right\rangle,
\end{eqnarray}
where $k_{\uparrow} = k$ and $k_{\downarrow} = [2m(E-\Delta)]^{1/2}/\hbar < k$.
The incident, transmitted, and reflected spin currents are
\begin{eqnarray}
{\bf Q}_{\rm in} &=& {\hbar^2 \over 2m{\Omega}} \left( k \sin(\theta) {\bf
\hat{x}}+ k \cos(\theta) {\bf \hat{z}} \right) \nonumber \\
{\bf Q}_{\rm trans} &=& {\hbar^2 \over 2m{\Omega}} \sin(\theta) k
\cos[(k_{\uparrow} - k_{\downarrow})x]
{\bf \hat{x}}
\nonumber\\
&&- {\hbar^2 \over 2m{\Omega}}
             \sin(\theta) k
\sin[(k_{\uparrow} - k_{\downarrow})x] {\bf \hat{y}}
\\
&&+ {\hbar^2 \over 2m{\Omega}} \left[
k \cos^2(\theta/2)- k_{\downarrow} \left({2k \over k + k_{\downarrow}}\right)^2
\sin^2(\theta/2)\right]
{\bf \hat{z}}
\nonumber \\
{\bf Q}_{\rm refl} &=& {\hbar^2 \over 2m{\Omega}} k \left( { k- k_{\downarrow}
\over k + k_{\downarrow} } \right)^2 \sin^2(\theta/2) {\bf \hat{z}}.  \nonumber
\end{eqnarray}
There are two points of physics that we wish to illustrate with this
example.  First, the transverse (perpendicular to ${\bf \hat{z}}$)
spin component of the reflected spin current
density is equal to zero. Since the total spin current density is
continuous at the interface, this means that all of the of the
transverse component of the incident spin current density is
transmitted through the interface; none is reflected.  In this toy
model, this result follows from our assumption that for spin up
electrons the height of the potential-energy step at the interface is
zero, so that the reflection amplitude for spin up electrons is zero
and the reflected part of the wavefunction is purely spin down.  For
models in which both components of spin experience a non-zero
potential-energy step, some of the incident ${\bf \hat{x}}$ component
of the spin current density will be reflected.  However, for many of
the materials combinations used commonly in metallic GMR devices,
like Cu/Co, Cu/Ni, or Cr/Fe, one of the spin components actually does
have a reflection amplitude close to zero over a large part of the
Fermi surface \cite{Stiles:1996,Xia:2002,Stiles:2002a}, so it is a
reasonable approximation in these cases that almost all of the
transverse (${\bf \hat{x}}$) component of the spin current density
will be transmitted into the ferromagnet.

The second important point of physics illustrated by the model
concerns what happens to the transverse component of the spin current
density after it enters the ferromagnet. The oscillatory ${\bf
\hat{x}}$ and ${\bf
\hat{y}}$ terms in the transmitted spin current density
represent precession of the spin about the ${\bf \hat{z}}$ axis
as a function of position as it penetrates through the magnet. In any
model in which there is a difference in exchange energy between
majority and minority spin states, the two spin components of a
wavefunction for a given eigenvalue $E$ must have different kinetic
energies, so that $k_{\uparrow} \ne k_{\downarrow}$ and the spin
state inside the magnet will precess.  The same phenomenon is
therefore present in more rigorous models. One can view this effect
as simply the
precession of the spin in the exchange field of the magnet. The period
of the precession, $2\pi /(k_{\uparrow} - k_{\downarrow})$ is very
short for a typical transition metal ferromagnet, on the scale of a
few atomic lattice spacings.  This is important because in real
3-dimensional
samples many electrons are incident on the magnetic layer from a
variety of directions, corresponding to states from all parts of the
Fermi surface, and therefore different
electrons take different paths through the magnetic layer.  Even if
all of the electrons begin with perfectly aligned spins at the
normal-metal/ferromagnet interface, electrons reaching a given depth
inside the magnet will have traveled different path lengths to get
there.  The result is classical dephasing.  Electron spins that have
traveled different path lengths will have precessed by different
angles around the ${\bf \hat{z}}$ direction, and therefore their ${\bf
\hat{x}}$ and ${\bf \hat{y}}$ components will not add constructively.
For locations more than a few atomic lattice constants into a magnetic
layer, when one sums over electrons from all relevant parts of the
Fermi surface in calculations that include first-principles
computations of the transmission amplitudes
\cite{Xia:2002,Stiles:2002a,Zwierzycki:2005}, the transverse (to
${\bf \hat{z}}$)
components of the spins average to zero.  As a consequence of this classical
dephasing, there is no net transmission of transverse spin angular
momentum through the ferromagnet.  The transverse angular momentum
that enters into the ferromagnet is effectively absorbed within a few
atomic layers from the interface.

For a full calculation of the spin torque at an interface in a real
3-dimensional sample, it is important to take into account not just
propagating wavefunctions, but also evanescent scattering states at
the interface.  If our toy model \#2 is generalized to three
dimensions, evanescent scattering wavefunctions are required in cases
where the incident electron approaches the interface from a glancing
angle, so that the part of the kinetic energy associated with the
perpendicular wavevector is less than the step height $\Delta$.  In
calculations with more realistic band structures, both evanescent and
propagating scattering states can couple to incident Bloch states
more generally, and it is necessary to take the evanescent states
into account to guarantee the continuity of the wavefunction and its
first derivative on the atomic scale.  Although evanescent scattering
states do not carry charge current, they do carry spin current, so
that they can contribute a significant spin torque even when the net
charge flow through the interface is zero.  This point can be
illustrated by our toy model if we consider a case in which the
spin-dependent step heights are sufficiently high that both of the
spin components are completely reflected.  The transmitted and
reflected parts of the scattering wavefunction are then
\begin{eqnarray}
\psi_{\rm trans} &=& {e^{-\kappa_{\uparrow} x} \over \sqrt{{\Omega}}}
\frac{2k}{k+i\kappa_{\uparrow}}
\cos(\theta/2) \left|
\uparrow \right\rangle
\nonumber\\
&&+ {e^{-\kappa_{\downarrow} x} \over \sqrt{{\Omega}}}
\frac{2k}{k+i\kappa_{\downarrow}}  \sin(\theta/2)
\left|\downarrow \right\rangle , \nonumber \\
\psi_{\rm refl} &=& {e^{-ikx}\over \sqrt{{\Omega}}}
\frac{k-i\kappa_{\uparrow}}{k+i\kappa_{\uparrow} }
\cos(\theta/2) \left|\uparrow \right\rangle
\nonumber\\
&&+ {e^{-ikx}\over \sqrt{{\Omega}}}
\frac{k-i\kappa_{\downarrow}}{k+i\kappa_{\downarrow} }
\sin(\theta/2) \left|\downarrow \right\rangle,
\end{eqnarray}
where $\kappa_{\uparrow}$ and $\kappa_{\downarrow}$ are decay
constants for the evanescent states in the ferromagnet.  Because the
transmission and reflection amplitudes are now complex, with (in
general) different complex phases for spin up and spin down
electrons, the transmitted and reflected spin current densities will
contain ${\bf\hat{y}}$ as well as ${\bf \hat{x}}$ components. The
transmitted spin current density also decays exponentially to zero as
a function of the penetration distance into the ferromagnet. What
this means is that, in effect, the electron penetrates into the
ferromagnet a distance on the order of $1/(\kappa_{\uparrow} +
\kappa_{\downarrow})$ and precesses around the exchange field as it
does so, so that when it emerges from the magnet it is rotated away
from its original orientation. Consequently, during the process of
reflection an electron can apply a torque to the magnet in both the
in-plane and perpendicular directions.

In calculations with realistic band structures for
normal-metal/ferromagnet interfaces, when one sums over the
Fermi surface to determine the total value of the transverse
part of the reflected spin current there is significant (but
not perfect) classical dephasing, so that the overall net flow of
reflected transverse angular momentum is close to zero
\cite{Stiles:2002a,Xia:2002,Zwierzycki:2005}. This means that the
incident transverse angular momentum that couples into the evanescent
states cannot end up flowing away from the interface through the
reflected states. Instead, this transverse spin angular momentum is
deposited in the interfacial region of the ferromagnet via the torque
from the evanescent states.
This contribution to the torque, which is not directly mediated by
the propagating states, can be described as due to spin filtering. If
one accounts for the evanescent states when constructing the
scattering states but
then ignores their contribution to the spin current, the
spin-filtering contribution corresponds to the resulting interfacial
discontinuity in the part of the spin current density carried by just
the propagating states.

The net result of the classical dephasing that occurs for both
transmitted and reflected electron waves at a
normal-metal/ferromagnet interface is that
the total transmitted and reflected spin currents, summed over all
relevant states on the Fermi surface, are approximately collinear
with the ferromagnetic layer's magnetization
(in the ${\bf \hat{z}}$ direction, in our toy model).  Since, to a
good approximation, no transverse angular momentum flows away from the
magnet, this collinearity means that approximately the {\it entire}
incident transverse spin current is absorbed near the
normal-metal/ferromagnet interface, and the spin transfer torque
(Eq.~(\ref{eq:pillbox})) becomes
\begin{eqnarray}
\label{eq:intabsorb}
               {\bf N}_{\rm st}
&=& A {\bf \hat{x}} \cdot ({\bf Q}_{\rm in} + {\bf Q}_{\rm refl} - {\bf Q}_{\rm
trans})
                          \approx A \hat{\bf x} \cdot{\bf Q}_{{\rm in}\perp}.
\end{eqnarray}

In terms of the parameters used in our first toy model above, it is
correct to say that when summing or averaging over all contributions
from around the Fermi surface that to a good approximation for a typical
metallic interface the dephasing leads
to $\langle {\rm Re}(t_{\uparrow}t_{\downarrow}^*)
\rangle = \langle {\rm Im}(t_{\uparrow}t_{\downarrow}^*) \rangle = \langle {\rm
Im}(r_{\uparrow}r_{\downarrow}^*) \rangle =0$,
and to a somewhat less-accurate approximation $\langle {\rm
Re}(r_{\uparrow}r_{\downarrow}^*) \rangle \approx 0$, so that on
average for our one electron
\begin{eqnarray}
                 {\bf N}_{\rm st}
\approx {A \over {\Omega}}{\hbar^2 k \over 2m} \sin(\theta) {\bf \hat{x}},
\label{eq:Model2}
\end{eqnarray}
and the spin torque acting on the magnet per unit area is equal to the
full component of incident spin current that is transverse to
the ferromagnet's moment.

The result in Eq.~(\ref{eq:intabsorb}) is a good approximation for
metallic interfaces, like Cr/Fe or Cu/Co, but the processes that lead
to the simple form $A \hat{\bf x} \cdot{\bf Q}_{{\rm in}\perp}$ may
be different for magnetic semiconductors or for tunnel junctions like
Fe/MgO/Fe.  Most electrons that scatter from tunnel barriers reflect,
whether they are majority or minority. In addition, tunneling is
dominated by electrons that are largely from particular parts of the
Fermi surface, so the classical dephasing processes that are important
for metallic junctions may be weaker in tunnel junctions. In fact,
there is good
evidence that $\langle {\rm Im}(r_{\uparrow}r_{\downarrow}^*) \rangle \ne 0$ in
tunnel junctions, so that for large applied biases there can be a
significant spin
torque component in the ${\bf \hat{y}}$ direction (perpendicular to the plane
defined by the incoming electron spin and the ferromagnet's moment)
\cite{Sankey:2007,Kubota:2007}.  This is discussed in more detail in
the article
by Sun and Ralph.

Before moving on to consider how the spin torque will affect the
ferromagnet's magnetization orientation, we wish to re-emphasize one last
important point.  Spin currents can flow within parts of devices even
where there is no net charge current.  Consequently, a spin transfer
torque can also be applied to magnetic elements that do not
carry any charge current \cite{Slonczewski:patent,Xia:2002}.  We have
already noted two examples of this effect, in Slonczewski's original
calculation of interlayer exchange coupling in a magnetic tunnel
junction \cite{Slonczewski:1989} and in our toy model \#2 for the
case when both spin-up and spin-down components of the wavefunction
are completely reflected.  Another important example occurs
in multiterminal normal-metal/ferromagnet devices, which are designed
so that a charge current flows only between two selected terminals,
but diffusive spin currents may also flow throughout the rest of the
device.  The groups of Johnson, van Wees, and others have demonstrated
non-local spin accumulation in nonmagnetic
wires using multiterminal devices
\cite{Johnson:1988,Jedema:2002,Tombros:2006,Lou:2007}.  Kimura et
al.\ have used a similar lateral device design to demonstrate
spin-torque-driven switching of a thin-film magnetic element which
carries no charge current \cite{Kimura:2006}.  One can view this
effect as due simply to a flow of spin-polarized electrons
penetrating by diffusive motion into a magnetic element and
transferring their transverse component of spin angular momentum,
while an equal number of electrons exit the magnet with an average
spin component collinear with the magnet so that they give no spin
torque.  In this way there can be a non-zero net spin current and
therefore a non-zero spin torque on the magnet, even when there is no net
charge current. Thus far the switching currents required in the
devices of Kimura {\it et al}.\ are larger than those needed to switch
comparable magnetic elements in standard magnetic-multilayer pillar
devices, because the magnitude of the spin current densities achieved
in the lateral devices (per unit injected charge current) is smaller
than in the multilayers.

{\em Spin Transfer Torque and the Landau-Lifshitz-Gilbert Equation.}
To calculate the effects of the spin transfer torque on magnetic
dynamics, in practice a term ${\dot{\bf M}}_{\rm st} \propto {\bf N}_{\rm
st}$ is generally simply inserted as an additional contribution on
the right side of the
Landau-Lifshitz-Gilbert equation (Eq.~(\ref{eq:llg})).  However, this
step deserves some careful consideration, as it involves a few subtle
points of physics. First, this insertion assumes that the all of the angular
momentum transferred from the transverse spin current density acts entirely to
reorient the orientation of the ferromagnet rather than, for example, being
absorbed in the excitation of short-wavelength magnon modes or being
transferred
directly to the atomic lattice.  This seems to be a reasonable
approximation for
describing the experiments performed to date; however, as we note below, the
existing spin-torque measurements in metallic multilayer samples are not
particularly quantitative.

A second simple matter to keep straight is the sign of
the torque.  An electron's magnetic moment is opposite to its spin
angular
momentum $\bm\mu = g_{e} \mu_{\rm B} {\bf S}/\hbar $, where ${\bf S}$
is the total spin and $g_{e} \approx - 2.0023$, and likewise in
transition metal ferromagnets the magnetization is generally opposite to the
spin density ${\bf M}=g \mu_{\rm B} {\bf s}/\hbar$, where ${\bf s}$ is
here the spin density and $g$ is typically in the range -2.1 to -2.2
\cite{Jonker:2004}.  We have defined ${\bf N}_{\rm st}$ as a time rate
of change of angular momentum.  Since the Landau-Lifshitz-Gilbert
equation is stated in terms of magnetization, the contribution of the
spin transfer torque should enter this equation with a sign opposite
to the change in angular momentum. In the end, the sign of the
spin transfer
torque is such as to rotate the spin angular momentum density of the
ferromagnet toward the direction of the spin of the incoming
electrons, or equivalently to rotate the magnetization of the
ferromagnet toward the direction of the moment of the
incoming electrons.

A third, potentially much more consequential, subtlety involves the orbital
contribution to the magnetization.  If we assume that all of the
angular momentum in the ferromagnet is due to its spin density, then
conservation of angular momentum implies that the effect of spin
transfer torque on the magnetization can be described simply by
inserting
\begin{eqnarray}
{\dot{\bf M}}_{\rm st} = -{\bf N}_{\rm st}
|g| \mu_{\rm B}/(\hbar \mathcal{V})
\label{eq:Mdot}
\end{eqnarray}
into the right side of the Landau-Lifshitz-Gilbert equation
(Eq.~(\ref{eq:llg})).
Here $\mathcal{V}$ is the volume of the ferromagnet (free layer) over
which the spin
torque ${\bf N}_{\rm st}$ is applied.
Ignoring any orbital contribution is a reasonable
first-order
approximation, because the orbital moments in transition metal
ferromagnets are largely quenched by the strong hybridization of the
d electrons.  However, spin-orbit coupling does give rise to a weak
orbital moment, typically less than a tenth of the spin moment as
indicated by the deviation of $g$ from -2. In a more precise treatment
that takes the orbital contribution into account, the total angular
momentum density would be ${\bf s}+\bm\ell$, and the magnetization
would be ${\bf M}= -\mu_{\rm B} (|g_{e}|{\bf s}+ \bm\ell)/\hbar$, so that
there might not be any simple proportionality between the total
angular momentum density and ${\bf M}$.  This
would necessitate a significantly more complicated picture, as
described in more detail immediately below.  However, in most
analyses of spin transfer torques, the potential effects of orbital
moments are ignored, and it is assumed that the spin torque is simply
described by Eq.~(\ref{eq:Mdot}).  The appropriateness of neglecting
orbital angular momentum is discussed briefly in the article by Haney,
Duine, N\'u\~nez, and MacDonald.

{\em A More Rigorous Approach to the Equation of Motion for
Magnetization.} In this section we will discuss how one might take a
more systematic approach to deriving the equation of motion for a
ferromagnet under the influence of a spin transfer torque.  This
exercise will give additional insights into what might be required to
account more accurately for effects like orbital angular momentum and
spin-orbit coupling.  This approach also provides a more natural
framework for considering spin transfer torques at domain walls and
in other spatially non-uniform magnetization distributions.

Our starting point is that the equation of motion for any
variable in quantum mechanics can be determined by taking the
commutator of the operator corresponding to that variable with the
Hamiltonian and then evaluating the
expectation value of the result.  To explore how this process works,
we will consider first the charge density, then the spin density, and
finally (briefly) the magnetization.  In second quantized notation
\cite{Baym} the
charge density and spin density operators are
\begin{eqnarray}
              \hat{n} &=& (-e)\sum_\sigma \hat\psi^\dagger_\sigma ({\bf r})
                                      \hat\psi_\sigma ({\bf r}) ,
\nonumber\\
              \hat{{\bf s}} &=& {\hbar\over 2} \sum_{\sigma,\sigma'}
                          \hat\psi^\dagger_\sigma ({\bf r})
                          \bm\sigma_{\sigma,\sigma'}
                          \hat\psi_{\sigma'} ({\bf r}) ,
\end{eqnarray}
in terms of the creation $\hat\psi^\dagger_\sigma$ and destruction
$\hat\psi_\sigma$ operators for an electron at point ${\bf r}$ and
spin $\sigma$.  These fermion operators obey the anti-commutation
relations $\{\hat\psi_\sigma({\bf r}),\hat\psi^\dagger_{\sigma'}({\bf
r}')\}=\delta({\bf r}-{\bf r}')\delta_{\sigma,\sigma'}$.  For the
present purposes we consider the Hamiltonian for non-interacting
electrons as in the mean field LSDA approach
\begin{eqnarray}
              \hat{{\mathcal H}} &=& {\hbar^2 \over 2m} \sum_\sigma \int d^3 r
              \bm\nabla \hat\psi^\dagger_\sigma ({\bf r})
              \cdot \bm\nabla \hat\psi_\sigma ({\bf r})
\nonumber\\
             &&+
             \int d^3 r \left[ V({\bf r}) \hat{n}({\bf r})
              + {2\mu_{\rm B} \over \hbar} {\bf B}_{\rm xc} \cdot \hat{{\bf
s}} \right]
\nonumber\\
             &&+ \int d^3 r
              \mu_0 {g\mu_{\rm B} \over \hbar}
              ( {\bf H}_{\rm ext} + {\bf H}_{\rm dip} ) \cdot \hat{{\bf s}} .
\end{eqnarray}
The first term in the above equation is the kinetic energy, the second
term is the potential, including the local exchange field, and the
third term is the coupling with the applied field ${\bf H}_{\rm ext}$
and the dipolar field ${\bf H}_{\rm dip}$ due to the rest of the
spins.  More generally, the potential, the local exchange field, and
the dipolar field are many-body terms, but for the present example
they are treated as effective single particle interactions.  For the
moment we are
also ignoring spin-orbit coupling in this Hamiltonian.

Let us first derive
the equation of motion for the electron charge density.  The only
term in the Hamiltonian that does not commute with the charge density
operator is the kinetic energy and it gives rise to a term that is the
divergence of the charge current density
\begin{eqnarray}
\label{eq:continuity}
               { d \hat{n} \over dt } &=& {1\over i\hbar}[ \hat{n},
   {\mathcal H} ]
\nonumber\\
&=& {-e\hbar \over 2im } \sum_\sigma \int d^3 r
\bigg\{
\hat\psi^\dagger_{\sigma'} ({\bf r}')
[\bm\nabla_{{\bf r}'}\delta(({\bf r}-{\bf r}'))]\delta_{\sigma,\sigma'}
\bm\nabla\hat\psi_\sigma ({\bf r})
\nonumber\\
&&-
\bm\nabla\hat\psi^\dagger_\sigma ({\bf r})
[\bm\nabla\delta(({\bf r}-{\bf r}'))]\delta_{\sigma,\sigma'}
\hat\psi_{\sigma'} ({\bf r}')
\bigg\}
\nonumber\\
&=& - \bm\nabla \cdot \hat{\bf j}.
\end{eqnarray}
The charge current density operator is
\begin{eqnarray}
              \hat{\bf j} &=& {-e\hbar \over 2im}
\sum_\sigma \left[
\hat\psi^\dagger_\sigma ({\bf r})
\bm\nabla\hat\psi_\sigma ({\bf r})
-\bm\nabla\hat\psi^\dagger_\sigma ({\bf r})
\hat\psi_\sigma ({\bf r})
\right] .
\end{eqnarray}
Taking the
expectation value of Eq.~(\ref{eq:continuity}) gives simply the
continuity equation for the charge density, as is required by charge
conservation.  The time rate of change
of the charge density in some volume is given by the net flux of
electrons into that volume.

Finding the time evolution of the spin density can be done using the
same method, but this exercise is somewhat more complicated because
the spin density is a vector and there are additional terms in the
Hamiltonian beside the kinetic energy with which it does not commute.
If we ignore spin-orbit coupling we get
\begin{eqnarray}
\label{eq:qdef}
               {d\hat{\bf s} \over dt} &=&
                - \bm\nabla \cdot \hat{\bf Q} -\gamma_0 \hat{\bf s}\times
               ({\bf H}_{\rm ext}+{\bf H}_{\rm dip}) ,
\end{eqnarray}
where $\hat{\bf Q}$ is the tensor spin current density operator
\begin{eqnarray}
              \hat{\bf Q} &=& {\hbar^2 \over 4im}
\sum_{\sigma,\sigma'} \bigg[
\hat\psi^\dagger_\sigma ({\bf r})
\bm\sigma_{\sigma,\sigma'}\otimes
\bm\nabla\hat\psi_{\sigma'} ({\bf r})
\nonumber\\
&&-\bm\nabla\hat\psi^\dagger_\sigma ({\bf r})
\otimes\bm\sigma_{\sigma,\sigma'}
\hat\psi_{\sigma'} ({\bf r})
\bigg] ,
\end{eqnarray}
and the dot product in Eq.~(\ref{eq:qdef}) connects to the spatial
index of the spin current.  Similar to the contribution in the
continuity equation, there is a contribution to the time rate of
change in the spin density due to the net flux of spins in and out of
a volume (the term involving $\hat{\bf Q}$).  In addition, the spin
density precesses in the local fields ${\bf H}_{\rm ext}+{\bf H}_{\rm
dip}$.  There is no contribution from the local exchange field because
(at least in the LSDA mean field theory) it is exactly aligned with
the expectation value of the local spin density.  The cross product
between these two quantities is then identically zero.

Comparing Eq.~(\ref{eq:qdef}) to Eqs.~(\ref{eq:mumagh}) and
(\ref{eq:llg}), the astute reader will notice some differences.
Eq.~(\ref{eq:qdef}) does not include a contribution from the
magnetocrystalline anisotropy, but this is just because for the
present we are ignoring the spin-orbit coupling.  More importantly,
Eq.~(\ref{eq:qdef}) appears not to include any term to account for
micromagnetic exchange.  The explanation for this is that there are
actually two contributions to the spin current density ${\bf Q}$ and
its divergence within a ferromagnet having a spatially non-uniform
magnetization.  The first is the non-equilibrium spin current that
flows with an applied bias and which is the contribution of interest
in this series of articles.  A second contribution is present in the
absence of any applied bias whenever the magnetization is
non-collinear.  This contribution can be viewed as the mediator of the
micromagnetic exchange interaction in analogy to Slonczewski's
calculation \cite{Slonczewski:1989} of the exchange coupling across a
tunnel barrier.  In general in the discussion of spin transfer
torques, and in particular in the rest of this article and the
accompanying articles, this contribution is taken into account by
including explicitly in the equation of motion a micromagnetic
exchange contribution in the form $- \bm\nabla \cdot {\bf Q}_{\rm eq}
=-\gamma_0 [A_{\rm ex}/(2\mu_0 M_{\rm s}^2)] {\bf s} \times \nabla^2
{\bf M}$, so that the spin current contribution then describes only
the non-equilibrium component.

We expect that the equations of motion become even more interesting
if one were to include spin-orbit coupling in the Hamiltonian.  First, there
are additional
terms in the equation of motion of the spin density,
Eq.~(\ref{eq:qdef}), because the contribution of spin-orbit coupling
to the Hamiltonian will not commute with the spin
density operator.  One new term generated by the spin-orbit coupling
is straightforward: the magnetocrystalline
anisotropy gives a contribution $-\gamma_0 {\bf \hat{s}}\times {\bf H}_{\rm
ani}$.  The damping term in Eq.~(\ref{eq:llg}) emerges as well, when a
coupling to a source of energy and angular momentum is also included.
The article by Tserkovnyak, Brataas, and Bauer describes the
derivation of such terms.  However, when the orbital angular momentum
in a ferromagnet is appreciable, one should recognize that the quantity of
primary interest in the Landau-Lifshitz-Gilbert equation is the
magnetization rather than the spin density, and these quantities need
no longer be simply proportional to each other.
The magnetization operator is $\hat{\bf M} = -\mu_{\rm B} (|g_{e}| {\bf
\hat{s}}+ {\bf \hat{\ell}})/\hbar$, where ${\bf \hat{\ell}}$ is the
orbital angular momentum density operator.  When taking the
commutator of $\hat{\bf M}$ with the Hamiltonian, the spin part of
the magnetization will generate the same divergence
of the spin current written in Eq.~(\ref{eq:qdef}) but there will be a
large number of additional terms due to spin-orbit coupling and ${\bf
\hat{\ell}}$.

These complications due to spin-orbit coupling
are likely to play an important role in the dynamics of
ferromagnetic semiconductors discussed in the article by Ohno and
Dietl, because spin-orbit coupling is much more significant in these
materials than in transition metal ferromagnets.
As the spin-orbit coupling is comparable to the exchange splitting
in the ferromagnetic semiconductors, the
band structure does not divide cleanly into majority and minority bands.  This
leads to additional complications in calculating transport properties
\cite{Nguyen:2007}, even beyond the complications
discussed above.
In these materials, it is not even clear that the spin current
is the most appropriate current to consider.  This question is related
to the issues of interest in the study of the spin Hall effect, see
\cite{Engel:2007} for a review.
Complications from spin-orbit coupling
are also likely to be amplified in ferrimagnetic samples as discussed
in the articles by Haney, Duine, N\'u\~nez, and MacDonald and by Sun
and Ralph.

\section{Multilayers and Tunnel Junctions}\label{sec:multi}

{\em Device Geometries.} For understanding the behavior of
spin-torque devices, the simplest
geometry to consider consists of two magnetic layers separated
by a thin non-magnetic spacer layer.  One magnetic layer serves to
spin-polarize a current flowing perpendicular to the layer interface
(this spin filtering can occur either in transmission or reflection),
and then this spin-polarized current can transfer angular momentum to
the other magnetic layer to excite magnetic dynamics.  The spacer
layer can either be a non-magnetic metal or a tunnel junction.  In
order that magnetic dynamics are excited in one magnetic layer but not
both, typically devices are designed to hold the magnetization in one
magnetic layer (the ``fixed'' or ``pinned'' layer) approximately
stationary at least for low currents.  This is done either by making
this layer much thicker than the other, so that it is more difficult
to excite by spin torque, or by fabricating it in contact with an
antiferromagnetic layer, which produces an effective field (``exchange
bias'') and increases the damping, which both act to keep the
ferromagnetic layer pinned in place.  The strength of the spin torque
acting on the thin ``free layer'' can be increased by sandwiching it
between two different pinned magnetic layers (with moments oriented
antiparallel), so that a spin-polarized current is incident onto the
free layer from both sides,
\cite{Berger:2003,Fuchs:2005,Huai:2005,Nakamura:2006,Meng:2006}.

Spin
transfer devices must be fabricated with relatively small lateral
cross sections, less than about 250 nm in diameter for typical
materials, in order that the spin torque effect dominates over the
Oersted field produced by the flowing current \cite{smallsize}.
(The Oersted field is often not negligible even in samples for which
the spin torque effect dominates -- it can be included in micromagnetic
simulations of the magnetic dynamics as an additional contribution to
${\bf H}_{\rm eff}$ in the Landau-Lifshitz-Gilbert equation as described
in the article by Berkov and Miltat.)
Small device
sizes are also convenient for understanding spin-torque-generated
magnetic dynamics because the macrospin approximation can become a
reasonable approximation, and of course small devices are desired
for many applications.

\begin{figure}
              \centering
              \resizebox{0.9\columnwidth}{!}{%
              \includegraphics{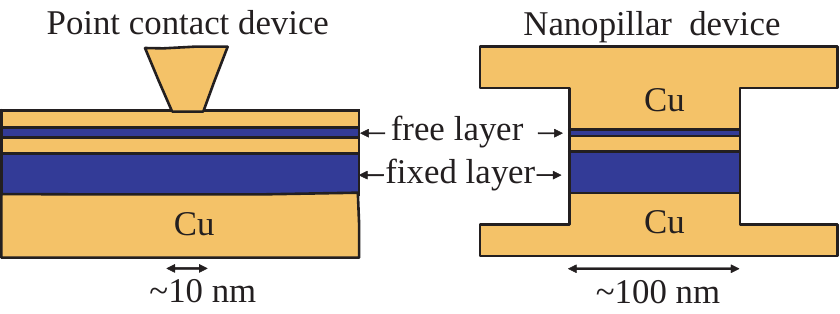}}
              \caption{Schematic experimental geometries.}
              \label{fig:geom}
\end{figure}

Two general experimental approaches are used to direct current
flow through an area $\le 250$~nm wide in a magnetic multilayer.  One
is to make electrical contact to an extended multilayer substrate in a
``point contact'' geometry, made either mechanically with a very sharp
tip \cite{Tsoi:1998} or using lithography techniques
\cite{Myers:1999,Rippard:2003}.  The other approach is to make
``nanopillar'' devices in which at least the free magnetic layer
and sometimes both the free and the fixed magnetic layers are patterned
to a desired cross section.  See Fig.~\ref{fig:geom}
for a schematic illustration
of both geometries.  Nanopillars can be made by electron-beam
lithography and ion milling \cite{Katine:2000}, stencil techniques
\cite{Sun:2002}, or electrodeposition of layers inside cylindrical
nanopores \cite{Wegrowe:2002}.  Samples in which the free layer is
patterned but the fixed layer is left as an extended film are
convenient for some experiments because this geometry minimizes the
magnetostatic coupling between the magnetic layers.  In general, point
contact devices require much more current density to excite magnetic
dynamics with spin torque, because the excitations must reorient a
small region in an extended magnetic film, working against strong
micromagnetic exchange.  Typical critical current densities for
excitations in point contact devices are $10^8$~A/cm$^2$ to
$10^9$~A/cm$^2$, while nanopillar devices have achieved current
densities of $< 10^7$ A/cm$^2$.  As far as we are aware, the record
low intrinsic (extrapolated to $T=0$) critical current in a
nanopillar device with a magnetic free layer that is thermally stable
at room temperature is 1.1 $\times$ 10$^6$ A/cm$^2$ \cite{Diao:2007}.
Two of the nice features of
nanopillars are that they can be made to have two different stable
magnetic configurations at zero applied magnetic field for memory
applications, and recently they have enabled direct x ray imaging of
current-driven magnetic dynamics \cite{Acremann:2006}.  Point contacts
have an advantage in that they give narrower linewidths ($\Delta f/f$)
as a function of frequency $f$ when used to
make spin-transfer nano-oscillators, as described below. Both point
contacts \cite{Myers:1999,Ji:2003,Chen:2004} and nanopillars
\cite{Ozyilmaz:2004} have also been used to study samples with a
single ferromagnetic layer.  The article by Katine and Fullerton
describes some of the strategies which can be employed to decrease the
critical current density for magnetic switching in nanopillars.  The
article by Silva and Rippard discusses research on point contact
devices.

{\em Methods for Calculating Spin Torques.} From a theoretical point
of view, a consensus has developed
\cite{Brataas:2006,Stiles:2006,Stiles:2007} for the general approach
to be used for calculating spin torques in metal multilayers.  We note
however, that there is not universal agreement that this model is
correct \cite{Levy,Urazhdin:2004,Wegrowe:2007}.
To solve for the spin transfer torque acting on a static magnetic
   configuration,
the consensus view is that one should determine the spin current
density ${\bf Q}$
through an analysis of the spin-dependent electron transport in the device
structure, and then identify the torque from the divergences of ${\bf Q}$
near magnetic interfaces or in regions of non-uniform magnetization.
(As we noted
above, this assumes that no angular momentum is lost to the excitation of
short-wavelength spin wave modes or to other processes.)

To calculate the spin torque on a moving magnetic configuration, a
somewhat more
sophisticated procedure is in principle necessary, for a rigorous treatment, in
order to take into account the ``spin-pumping'' effect \cite{Tserkovnyak:2002}.
Spin pumping refers to the fact that a precessing magnetization can produce a
non-zero spin current density in a magnetic device, in addition to the spin
current density generated directly by any applied bias.
The total spin torque on the moving magnetization should be determined from the
divergence of the total spin current arising from both sources.
However, typically the primary effect of spin pumping is merely to increase the
effective damping of the ferromagnet in a way that is independent of bias
(although it may depend strongly on the precession angle
\cite{Tserkovnyak:2003}),
so that spin pumping can be taken into account approximately just by
renormalizing
the damping constant.  Then the spin transfer torque even on a moving
magnetization configuration can be determined, approximately, from
the divergence
of just the bias-dependent part of the spin current density.

Once the spin torque is calculated, the response of the magnetization
is generally
determined by inserting this torque as an additional contribution to
the classical
Landau-Lifshitz Gilbert equation
of motion (Eq.~(\ref{eq:llg})).
The article by Haney,
Duine, N\'u\~nez, and MacDonald describes this consensus approach as
the ``bookkeeping theory.''
Spin pumping can also be incorporated into fully
self-consistent solutions without too much extra difficulty, because
spin pumping
is local in time, depending only on ${\bf M}$ and ${\dot{\bf M}}$,
and it can be
calculated within the same type of theoretical framework needed for calculating
the direct bias-generated spin torque \cite{Tserkovnyak:2003,Polianski:2004}.

Depending on the amount of disorder in the device and other details,
there are a number of strategies for computing the spin
current density, some of which are compared in Ref.~\cite{Bauer:2001}.
If scattering within a layer is weak enough that most electrons do not
scatter except at interfaces, the transport is called ballistic. The opposite
limit, in which most electrons scatter
several times while traversing a layer, is called diffusive.
Calculations for particular devices can be complicated by the fact
that some layers may be in the diffusive limit while others are in the
ballistic limit and some layers are in between.

The ballistic regime can be
treated by constructing the scattering
states of the system \cite{Waintal:2000,Zwierzycki:2005},
by the Keldysh formalism \cite{Edwards:2004} or by non-equilibrium
Green's functions \cite{Haney:2007}.
References \cite{Edwards:2004} and
\cite{Haney:2007}  have considered some of the potential effects of quantum
coherence in the ballistic limit, but these effects are generally not
expected to
be important in the types of metallic multilayer samples typically studied
experimentally, because their interfaces are not sufficiently abrupt
and perfect.
In principle, scattering-state formalisms, the Keldysh method, and
non-equilibrium
Green's functions techniques can all be extended into the diffusive regime, but
the strong scattering in the diffusive regime is generally easier to
treat using
the semiclassical approaches discussed next.

The Boltzmann
equation \cite{Stiles:2002b,Shpiro:2003}  neglects coherent effects,
but is accurate for both ballistic and diffusive samples, and can
interpolate between these limits.  In the Boltzmann-equation
approach, the transport is described in terms of a semiclassical
distribution function so that the behavior of electrons on different
parts of the Fermi surface are tracked separately.
Other
calculation strategies are based on approximations that sum over this
distribution function and compute the transport in terms of just its
moments.  These include the
drift-diffusion approximation
\cite{Berger:1998,Grollier:2001,Stiles:2004} and circuit theory
\cite{Brataas:2000}.
For the non-collinear magnetic configurations
considered here, these approximations are extensions of
Valet-Fert theory \cite{Valet:1993}, which is widely used to describe
GMR for collinear magnetizations.

In all of these approaches, the transport across the interfaces is
described in terms of spin-dependent transmission and reflection
amplitudes.  For electron spins that are collinear with the
magnetization, these processes are incorporated into the transport
calculations as boundary conditions between the solutions of the
transport equations in each layer, and the
results are spin-dependent interface resistances
\cite{interfaceresistance} (or conductances).  For spins that are not
collinear, transmission and reflection give rise both to boundary
conditions on the transport calculations and also give the spin
transfer torque.  In the drift-diffusion approach, the non-collinear
boundary conditions are that the transverse spin current is
proportional to the transverse spin accumulation
\begin{eqnarray}
              {\bf Q}_\perp^{\rm NM} \cdot \hat{\bf n}= w {\bf
   m}_\perp^{\rm NM} ,
\end{eqnarray}
where $\hat{\bf n}$ is the interface normal, $w$ is a characteristic
velocity, ${\bf m}_\perp^{\rm NM}$ is the transverse spin
accumulation, and the ${\rm NM}$ superscript indicates that the
transverse spin density and spin current are evaluated in the
non-magnet as they are both zero in the ferromagnet.  This boundary
condition has a straightforward interpretation.  Since the interface
acts as an absorber of any transverse spin component that scatters
from it, there is no out-going spin current to cancel the incoming
spin current, so there must be a net accumulation of transverse spins
in the non-magnet near the interface.  Since the incident transverse
spin current is equal to the spin transfer torque it is also the case that the
transverse spin accumulation is proportional to the spin transfer
torque.  The relation leads some authors to discuss separate spin
current and spin accumulation mechanisms for the spin transfer torque.
However, from this discussion, we see that the two are intimately
related, and the spin transfer torque can be fully accounted for in
terms of the spin current; there is not an extra separate contribution from
spin accumulation.

For the case of a symmetric two-magnetic-layer device with a metal
spacer, Slonczewski \cite{Slonczewski:2002} calculated the spin
transfer torque using a simplified Boltzmann equation grafted with
circuit theory.  He found that the torque on the free layer
magnetization ${\bf M}$ due to the misalignment with fixed layer
magnetization ${\bf M}_{\rm fixed}$ can be described by adding
to the Landau-Lifshitz-Gilbert equation (Eq.~(\ref{eq:llg})) a term
of the form
\begin{eqnarray}
                 {\dot{\bf M}}_{\rm st}
= \eta(\theta){\mu_{\rm B} I\over e \mathcal{V}}~
                 \hat{\bf M}\times(\hat{\bf M}\times\hat{\bf M}_{\rm fixed}),
                 \label{eq:Nst}
\end{eqnarray}
where $I$ is the current, $\mathcal{V}$ is the free-layer
volume on which the spin torque acts,  $\eta(\theta) = {q/( A+B\cos\theta)}$,
$\hat{\bf M}$ and $\hat{\bf M}_{\rm fixed}$ are unit vectors in the
directions of ${\bf M}$ and ${\bf M}_{\rm fixed}$ (not operators), and
$\cos\theta=\hat{\bf M}\cdot\hat{\bf M}_{\rm fixed}$.  All of the
details of the layer structure are buried in the constants $q$, $A$,
and $B$. Very similar results have also been found by a variety of
other theoretical approaches.  Below, we will sometimes refer loosely to
${\dot{\bf M}}_{\rm st}$ as a ``torque'', even though strictly speaking
its units are {\it (magnetic moment)/(volume} $\cdot$ {\it time)} rather
than {\it (angular momentum)/time}.

We note that the direction of ${\dot{\bf M}}_{\rm st}$ indicated by
Eq.~(\ref{eq:Nst})
is exactly what is expected from the simple picture of
Eq.~(\ref{eq:intabsorb}), based on the approximately complete absorption of the
transverse spin current by the magnetic free layer.  When the current
has the sign that electrons flow from the fixed layer to the free
layer in a multilayer like Co/Cu/Co,
the electron spin moment incident on the free layer is in the same
direction as ${\bf M}_{\rm fixed}$ and the double cross product
in Eq.~(\ref{eq:Nst})
represents just the transverse component.  When the current is
reversed, it is the electrons reflected from the fixed layer that
apply a torque to the free layer; their moments are on average oriented
antiparallel to ${\bf M}_{\rm fixed}$ on account of the reflection,
and therefore the transverse component of spin current incident on the
free layer changes sign.  Subsequent calculations
\cite{Manschot:2004a,Xiao:2004} have generalized Eq.~(\ref{eq:Nst}) for
asymmetric structures.  Ref.~\cite{Xiao:2007} compares these simple
forms to full calculations using the Boltzmann equation and shows that
they agree for typical layer thicknesses but break down in some
limits.  That paper also shows where a drift-diffusion approximation
fails to reproduce the results of the Boltzmann equation calculations.

\begin{figure}
              \centering
              \resizebox{0.6\columnwidth}{!}{%
              \includegraphics{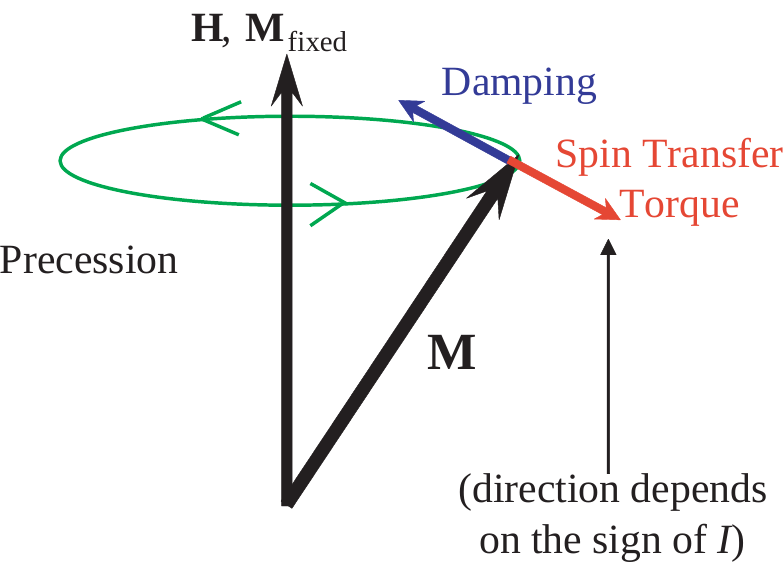}}
              \caption{Directions of damping and spin-torque vectors for a
              simple model discussed in the text.}
              \label{fig:precess}
\end{figure}

\begin{figure*}
              \centering
{%
              \includegraphics{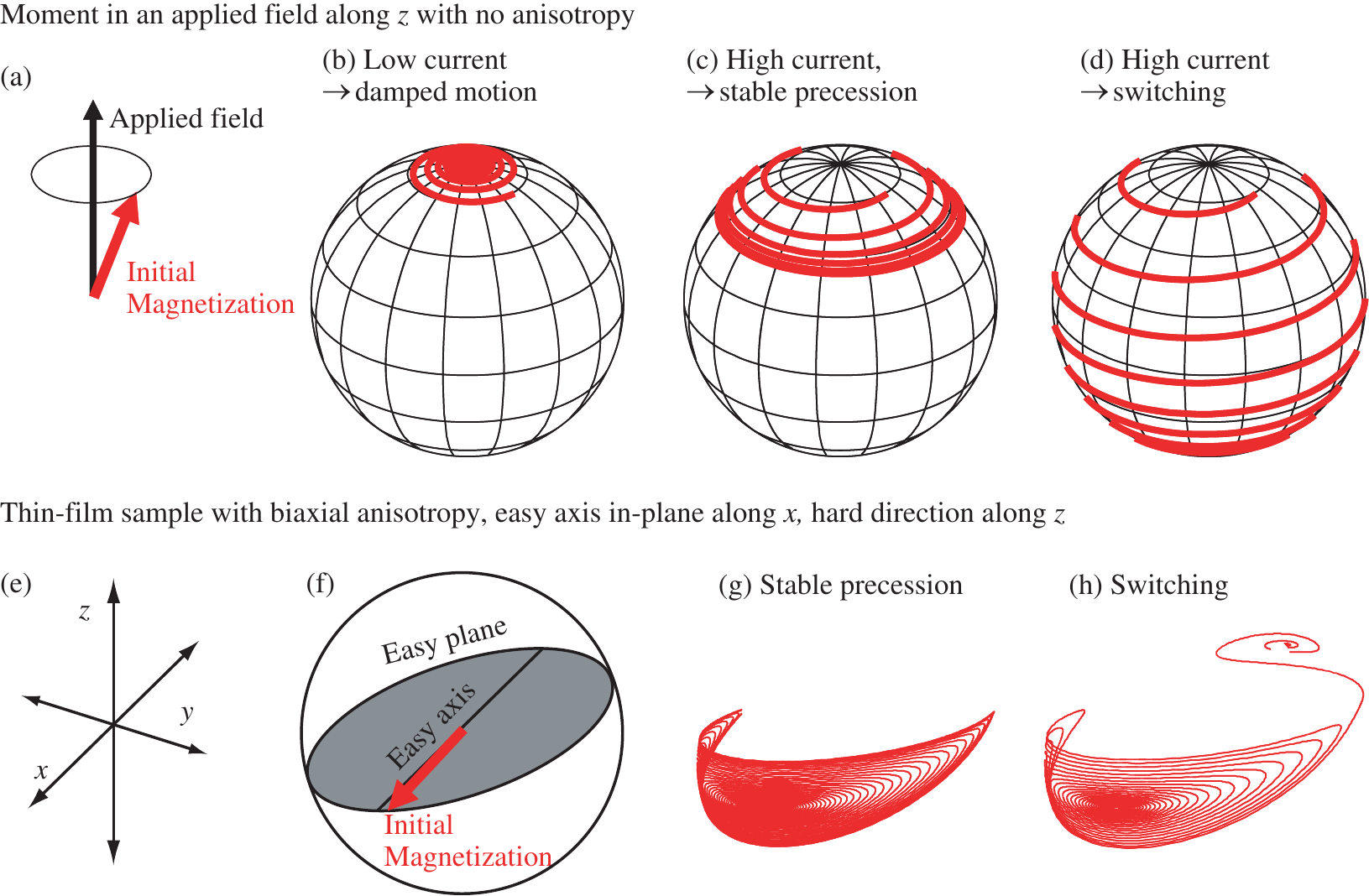}}
\caption{Trajectories of spin-torque-driven dynamics for the
       magnetization vector ${\bf M}$. (a) Initial magnetic configuration
assumed for panels (b,c,d), with the free layer magnetization slightly
misaligned from $\hat{\bf z}$, for example due to a thermal
fluctuation.  For these panels, we assume a sample with no magnetic anisotropy,
with ${\bf M}_{\rm fixed}$ and the applied magnetic field aligned in
the $\hat{\bf z}$ direction.
(b) For currents below a critical current, ${\bf M}$ spirals back toward
the low-energy $\hat{\bf z}$ direction on account of magnetic damping.
(c) For currents larger than the critical value, the spin transfer torque
causes the effective damping to become negative, meaning that ${\bf M}$
spirals away from $\hat{\bf z}$, with a steadily-increasing precession
angle.  The ultimate result can be either stable steady-state precession
at large precession angle (shown in (c)) or magnetic reversal (shown in
(d)), depending on angular dependence of the spin torque and damping.
(e,f) Geometry assumed for a thin-film magnetic sample with a strong
easy-plane anisotropy and a weaker uniaxial anisotropy with stable static
magnetic states along the $\pm \hat{\bf x}$ directions.  For this
geometry, the spin transfer torque from a direct current can also
produce either (g) steady-state precession or (h) magnetic reversal
from $+ \hat{\bf x}$ to $- \hat{\bf x}$.
}
              \label{fig:traject}
\end{figure*}

{\em Spin-Transfer-Driven Magnetic Dynamics.}
The qualitative types of magnetic dynamics that can be excited by
${\dot{\bf M}}_{\rm st}$ with the form given by Eq.~(\ref{eq:Nst})
can be understood using the diagrams shown in Fig.~\ref{fig:precess}
and Fig.~\ref{fig:traject}.
We first consider
the simplest possible geometry, in which the free layer magnetization
${\bf M}$ is assumed to move as one macrospin and the magnetic-field
direction and the fixed layer moment ${\bf M}_{\rm fixed}$ both point
along ${\bf
\hat{z}}$.  We also assume, initially, that there is no magnetic
anisotropy.
We will consider the problem in terms of a linear stability analysis, to
see, when the free layer moment is initially perturbed slightly from
the field direction, whether it returns to rest or whether a
current can destabilize it to generate large-angle dynamics.
This analysis can be achieved using
the Landau-Lifshitz-Gilbert equation of motion (Eq.~(\ref{eq:llg})).
In the absence of any spin transfer torque or damping, if the free
layer moment ${\bf M}$ is
instantaneously tilted away from ${\bf \hat{z}}$ then it will precess
in a circle, due to the torque from the applied magnetic field.
(For real thin-film devices, magnetic anisotropies generally
cause the precessional motion to trace out an ellipse.) If there
is damping in addition to an applied magnetic field (but still no
applied current),
the torque due to damping will push ${\bf M}$
back toward the low-energy configuration along ${\bf \hat{z}}$.
Consequently, if ${\bf
M}$ is perturbed away from ${\bf \hat{z}}$, then at $I=0$ it will precess
with gradually decreasing precession angle back
toward ${\bf \hat{z}}$ along a spiral path.  This type of magnetic
trajectory is depicted in Fig.~\ref{fig:traject}(b).

When a current is applied, the direction of the spin transfer torque
predicted by Eq.~(\ref{eq:Nst}) is either parallel to the damping
torque or antiparallel to it, depending on the sign of the current
(see Fig.~\ref{fig:precess}).
(For the more realistic case of elliptical precession in the presence
of magnetic anisotropies, the
instantaneous orientations of the spin torque and the damping are not
always collinear, but on average over each cycle the spin torque can
still be understood as either reinforcing or acting opposite to the
damping.)  For the sign of the current that produces a spin-torque
contribution ${\dot{\bf M}}_{\rm st}$ in
the same direction as the damping, there are no current-induced
instabilities in the free-layer orientation.  The current increases
the value of the effective damping, and ${\bf M}$ simply spirals more
rapidly back to the ${\bf \hat{z}}$ direction after any perturbation.
For small currents of the opposite sign, such that
${\dot{\bf M}}_{\rm st}$ is
opposite to the damping but weaker in magnitude, the spin torque just
decreases the effective damping, and again nothing exciting happens, at
least at zero temperature.  (Effects of thermal fluctuations are discussed
below.)

When ${\dot{\bf M}}_{\rm st}$ is opposite to the damping torque
and of
greater magnitude, then following any small perturbation ${\bf M}$
will spiral away from the low-energy configuration along
${\bf \hat{z}}$ to
increasing angles -- the current destabilizes the orientation with
${\bf M}$ parallel to ${\bf M}_{\rm fixed}$
and may excite large-angle dynamics.  In effect, a sufficiently large
current drives the damping to be negative, which leads to
the amplification of any deviations of ${\bf M}$ from equilibrium.
Past this point of instability, the large-angle dynamics excited by
spin transfer can fall in two broad classes within the macrospin
approximation, depending on the angular dependencies of the spin
transfer torque, the damping torque, and the
magnetic anisotropy.
One possibility, which may occur if the damping torque increases
with precession angle faster than the spin torque, is that the initial
increase in precession angle may eventually be limited, so that
${\bf M}$
may achieve a state of dynamical equilibrium,
precessing continuously at some fixed average angle in response
to the direct current
(see Fig.~\ref{fig:traject}(c)). In this state, the
energy gained from the spin torque during each cycle of precession is
balanced by the energy lost to damping.  The second possible class of
spin-torque-driven magnetic dynamics is that the precession angle may
be excited to ever-increasing values until eventually it reaches
180$^{\circ}$, meaning that ${\bf M}$ is reversed
(see Fig.~\ref{fig:traject}(d)).

In real samples, the magnetic anisotropy can generally not be ignored,
as we have done so far in the discussion above.  The
typical sample formed as part of a thin-film magnetic multilayer will
have biaxial magnetic anisotropy, consisting of a strong
in-plane anisotropy and a weaker uniaxial anisotropy along one
of the in-plane axes.  In Fig.~\ref{fig:traject}(e)-(g) we consider
some of
the magnetic trajectories that can be excited by spin transfer for
this type of
sample, where the easy magnetic axes are assumed to lie along the
$\pm \hat{\bf x}$ directions, and ${\bf M}_{\rm fixed}$ and the
applied magnetic field are also assumed to point along
$\hat{\bf x}$ instead of the $\hat{\bf z}$ direction as
in Fig.~\ref{fig:traject}(a)-(d).
Within the macrospin approximation, the
dynamics of ${\bf M}$ excited by the spin transfer torque can still
be calculated by integrating the
Landau-Lifshitz-Gilbert equation
(Eq.~(\ref{eq:llg})) with the spin-transfer-torque
term Eq.~(\ref{eq:Nst})
included.

Different behavior occurs in different regimes.
Consider first in-plane magnetic fields less
than the coercive field needed to produce
magnetic-field-induced switching.  At low temperature, as the current
increases, there is generally first a critical current which leads to
states of dynamical equilibrium in which
${\bf M}$ undergoes steady-state precession along an
approximately elliptical trajectory
(see Fig.~\ref{fig:traject}(g))
\cite{Kiselev:2003,Krivorotov:2004,Devolder:2005}. For
slightly larger currents, the
precessing state becomes unstable, and the precession angle for
${\bf M}$ increases until it reaches
180$^{\circ}$, thereby achieving switching to the state with ${\bf
M}$ antiparallel to ${\bf
M}_{\rm fixed}$ (Fig.~\ref{fig:traject}(h)). For applied magnetic fields
larger than the coercive
field, usually only steady-state
precessional dynamics are observed (Fig.~\ref{fig:traject}(g)).
Additional static and dynamic magnetic states may occur at even larger
values of current (see below).
If one starts
with the free-layer moment ${\bf M}$ oriented antiparallel to the fixed layer,
rather than parallel ({\it i.e.}, with ${\bf M}$ in the
$-\hat{\bf x}$ direction rather than $+\hat{\bf x}$), a reversed
sign of current is required to
produce a negative effective damping, and this can excite
steady-state dynamics or switch ${\bf M}$ back to the parallel
orientation.

\begin{figure}
              \centering
              \resizebox{0.9\columnwidth}{!}{%
              \includegraphics{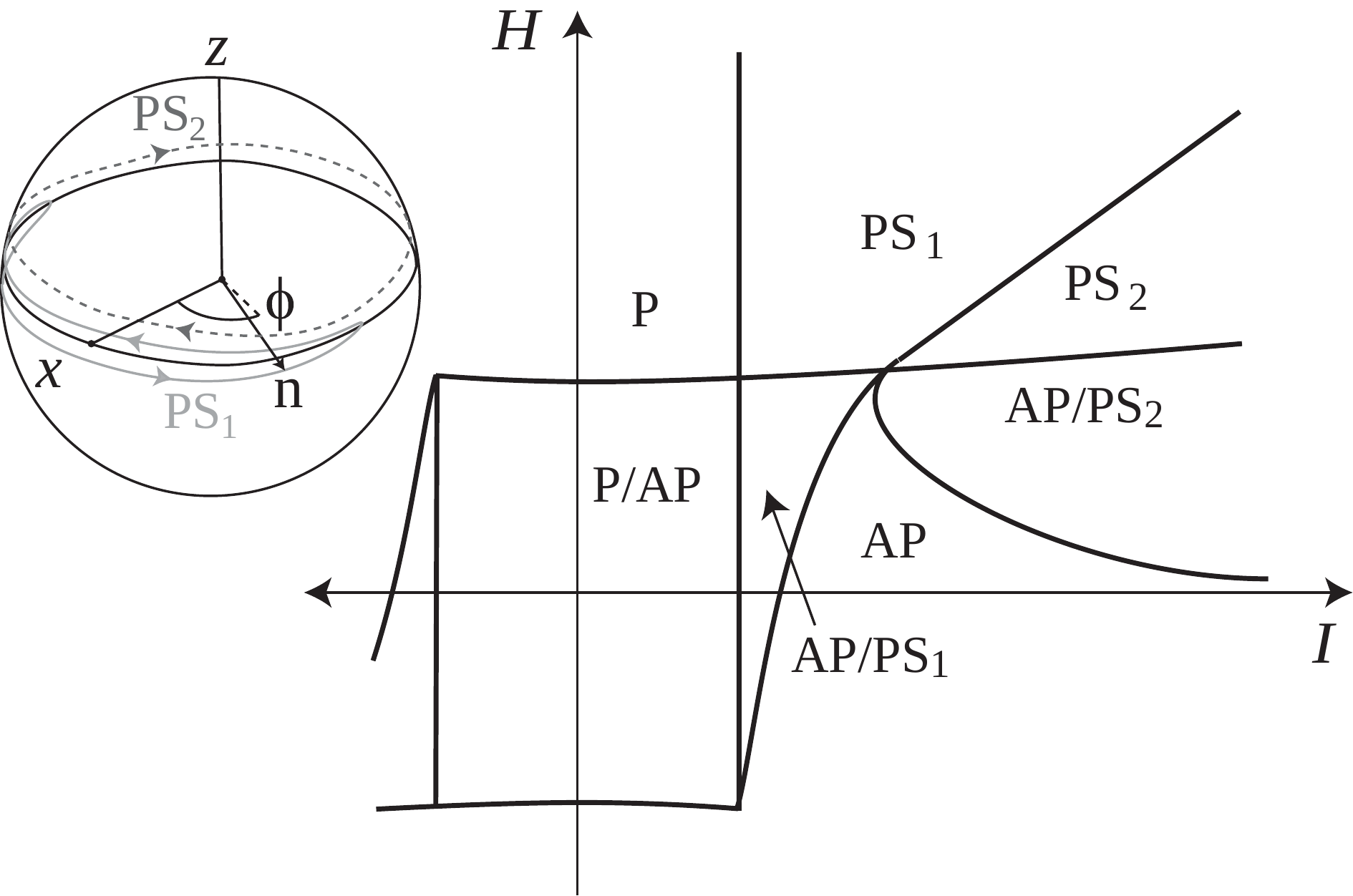}}
\caption{Schematic phase diagram for spin-torque-driven magnetic
       dynamics of a thin-film free layer in a nanopillar device at zero
       temperature as a function of current $I$ and a magnetic field $H$
       applied in-plane along with magnetic easy axis.  P is the static
       state with the free-layer magnetization parallel to the fixed-layer
       magnetization, AP is the static antiparallel state, and PS$_{1}$
       and PS$_{2}$ correspond to the steady-state precession states
       depicted in the inset.  Where two states A and B are indicated
       using the notation A/B, both states are stable at zero temperature.
       Diagram courtesy of Yaroslaw Bazaliy \cite{Bazaliy:2007}.
       }
              \label{fig:pd}
\end{figure}

For thin-film samples with the magnetizations of both the fixed
and free layers oriented in plane and with their easy axes aligned
with the applied magnetic field, it is possible to estimate
the threshold current for small-angle
excitations in a macrospin picture.  The estimation proceeds by integrating
over the elliptic orbit for a small precession angle to determine both
the energy lost due to damping and the energy gained from the spin
torque during each cycle, and determining at what value of current the
spin torque overcomes the damping.  This leads, for example, to the
critical current
expression for excitations from an initially parallel magnetic
orientation ($\theta = 0$)
\begin{eqnarray}
I_c = {2e \over \hbar} {\alpha \over \eta(0)} \mathcal{V} \mu_0 M_s \left(H +
H_k + {M_s \over 2 } \right),
\label{eq:critI}
\end{eqnarray}
referred to by the article of Katine and Fullerton.  Here
$\alpha$ is the Gilbert damping,
$H$ is the applied magnetic field and
$H_k$ the strength of the within-plane magnetic anisotropy
\cite{Katine:2000,Sun:2000}.  The saturation magnetization $M_{\rm s}$
appears in the last factor because of the large thin-film
demagnetization effect which favors an in-plane orientation for ${\bf
M}$; $M_{\rm s}/2$ is typically much larger than $H$ or $H_k$.  It is
important to note that this threshold describes only the first
instability of the free layer to small angle precession. There is a
separate threshold for magnetic switching at slightly larger currents,
which has a different dependence on magnetic field.
Approximate analytical expressions been derived to
describe the switching threshold and other boundaries between the
different dynamical states that can be driven by spin torques
\cite{Bertotti:2005,Stiles:2006,Bazaliy:2007}.  These phase boundaries
are illustrated in Fig.~\ref{fig:pd} for the case of a magnetic field applied
in-plane along the easy axis of the magnetic free layer.

One common result of all semiclassical calculations of transport and
spin transfer torque for metallic multilayers
\cite{Brataas:2006,Stiles:2006,Stiles:2007}
is that they predict a fairly substantial asymmetry between the
differential torque near parallel and antiparallel alignment.  That is,
\begin{eqnarray}
              \left| { d {\dot{\bf M}}_{\rm st} (\theta) \over d \theta
              }_{\theta=\pi}\right|
              >
              \left| { d {\dot{\bf M}}_{\rm st} (\theta) \over d \theta
              }_{\theta=0}\right| .
\end{eqnarray}
In other words, $B$ is comparable to $A$ in $\eta(\theta) =
{q/(A+B\cos\theta)}$.  This asymmetry arises from the different
amounts of spin accumulation for alignments close to parallel as
compared to those close to antiparallel.  The results suggest that the
critical currents for switching out of the antiparallel state
in metallic multilayers should
be significantly less than those for switching out of the parallel
state.  However, typically the asymmetry found in the measured
critical currents is much less than
expected \cite{Braganca:2005}.  The degree of asymmetry seems to
depend on the sample structure ({\it e.g.}, on the amount of taper in
nanopillar sidewalls), so the discrepancy could result in part from
deviations from a completely uniform magnetization state, which might
affect either the mechanism of reversal or the degree of spin
accumulation. On the
other hand, measurements of noise instabilities for magnetizations
close to perpendicular \cite{Smith:2006} suggest an asymmetry
consistent with the theoretical expectations.

If ${\bf M}_{\rm fixed}$ and ${\bf M}$ are
misaligned within the sample plane by an applied magnetic field or an
exchange bias, the critical currents will depend on the angle
$\theta_0$ between the equilibrium orientations of ${\bf M}$ and ${\bf
M}_{\rm fixed}$.  This can be understood in a simple way by
Taylor-expanding the instantaneous value of the spin torque in the sample
plane: ${\dot{\bf M}}_{\rm st} (\theta) \approx {\dot{\bf M}}_{\rm
st} (\theta_0)+ [{d\dot{\bf M}}_{\rm st} (\theta_0)/d\theta] (\theta -
\theta_0)$.
For current pulses much larger than the threshold needed to produce
excitations,
the first term can
dominate the fast magnetic dynamics that are produced.  However, for
currents near the critical current this term will generally just cause the
equilibrium angle to shift by a small amount
as a function of $I$, because it can be opposed by strong anisotropy
forces.  Near
the excitation threshold, it is actually
the second term of the Taylor expansion which governs the
stability of the static state, because it determines whether the spin
torque has the sign to increase or decrease the magnitude of
deviations from the equilibrium angle.  Dynamical excitations will
occur at zero temperature when $d{\dot{\bf M}}_{\rm st}
(\theta_0)/d\theta$ has the correct
sign and
becomes sufficiently large to overcome the intrinsic damping, so as
to cause the
free-layer moment to spiral away from its equilibrium orientation
with increasing
precession angle, as discussed above.
In a macrospin picture $I_c \propto 1/[d{\dot{\bf M}}_{\rm st}
(\theta_0) /d\theta]
\approx 1/\cos(\theta_0)$ \cite{Sun:2000,Mancoff:2003}.

An additional subtlety that can complicate understanding
spin-torque-driven magnetic dynamics is that the magnetization in
ferromagnetic devices is never completely spatially uniform, but even
in thin-film devices with very small cross section there is some
spatial dependence to the magnetization within the layer.  This has
the consequence that even in devices with electric current flowing
strictly perpendicular to the plane the spin current density will have
a non-zero spatial component within the plane.  These lateral spin
currents can produce additional instabilities that lead to the
excitation of spatially non-uniform magnetization dynamics within a
magnetic layer \cite{Polianski:2004,Stiles:2004,Adam:2006}, and in
fact these excitations have been observed in devices containing only a
single magnetic layer \cite{Ozyilmaz:2004} and standard
two-magnetic-layer devices \cite{Ozyilmaz:2005,Sankey:2006}.  Magnetic
excitations that are non-uniform through the thickness of magnetic
layers may also be possible \cite{Myers:1999,Ji:2003}.  Initial
attempts to incorporate lateral spin transport self-consistently with
spatially dependent magnetic dynamics in micromagnetic calculations of
multilayer devices are under way.

{\em Measurements of Magnetic Dynamics.}
The main experimental probe of the spin-torque-driven magnetization
dynamics is measuring the resistance.  The resistance reflects the
magnetic configuration through the GMR effect.  One signature that a
current-driven change in magnetic configuration is due to the spin
transfer effect is that it is asymmetric in the direction of the
current, for the reasons explained above.  In the first identification
of a spin transfer effect \cite{Tsoi:1998}, Tsoi {\it et al}.\ measured a
peak in the differential resistance for one direction of the current
and not the other.  In switching devices
\cite{Myers:1999,Katine:2000}, for large currents flowing from the
fixed layer to the free layer (electrons flowing from free to fixed),
the free layer magnetization is driven to be antiparallel
to the fixed layer,
resulting in
the high resistance state, while the opposite current drives the
sample to the parallel, low-resistance state.  The sign of the torque
observed experimentally agrees with the sign predicted by theory.

\begin{figure}
              \centering
              \resizebox{0.6\columnwidth}{!}{%
              \includegraphics{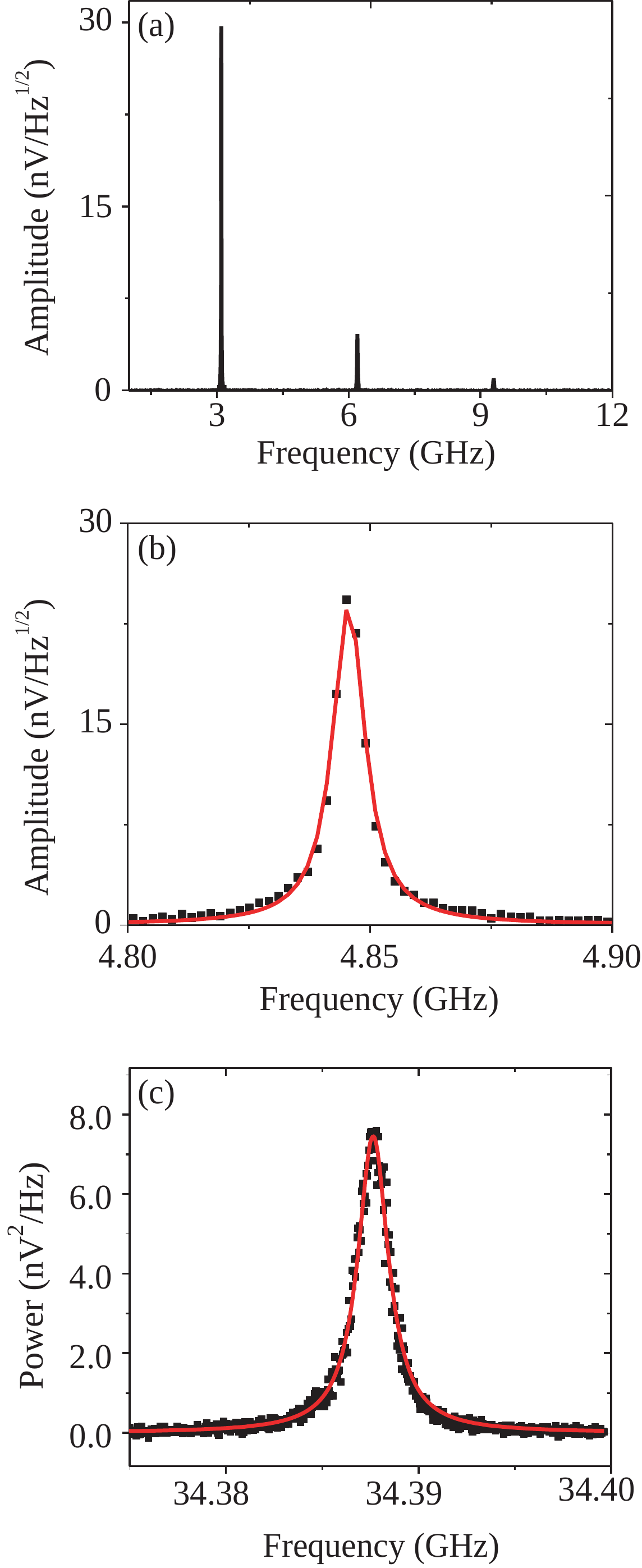}}
\caption{Spectra of voltage oscillations produced by
       spin-torque-driven steady-state magnetic precession, showing
       several harmonics and the linewidths of the signals.  (a,b) Spectra
       for all-metal nanopillar samples taken at a temperature of 40 K,
       courtesy of Ilya Krivorotov.  (c) Linewidth for an all-metal point
       contact sample at room temperature from the NIST, Boulder group
       \cite{Rippard:2004b}.
       }
              \label{fig:spectra}
\end{figure}

A sample excited into a state of dc-driven magnetic precession
naturally emits a substantial microwave-frequency signal at its
electrical contacts.  The resistance is changing at microwave
frequencies and a dc current is applied, so by Ohm's law a microwave
voltage is generated.  This can be detected either in the frequency
regime \cite{Kiselev:2003,Rippard:2004} or directly in time-domain
measurements \cite{Krivorotov:2005,Krivorotov:2007}.  This mode of
steady-state precession is of interest for applications which might
benefit from a
nanoscale oscillator or microwave source that can have a narrow
linewidth and is tunable in frequency, see Fig.~\ref{fig:spectra}.

The phase diagram of
the spin-transfer-driven dynamics as a function of current, magnetic
field, and the direction of magnetic field contains several distinct
types of precessional modes and static magnetic states
\cite{Kiselev:2003,Krivorotov:2007a}.  When one crosses from one mode
to another by varying current or field, the frequency of precession
can jump, the microwave power output can increase or decrease dramatically,
the linewidth can change, and the resistance measured by standard
low-frequency techniques generally changes by a small amount, too.
When a device is biased near the boundary between two different modes,
it often exhibits time-dependent switching behavior between them
\cite{Urazhdin:2003,Krivorotov:2004,Krivorotov:2007}. The dynamical
phase diagram has
been mapped in some detail for both in-plane and perpendicular
magnetic fields in nanopillar devices
\cite{Kiselev:2003,Bertotti:2005,Ozyilmaz:2005,Ozyilmaz:2003,Li:2003,Covington:2004,Bazaliy:2004,Kiselev:2004,Xiao:2005},
and it shows surprisingly good agreement even with the simplest
macrospin models, with some exceptions at large currents and for
devices in which the magnetic configuration begins in a vortex state
\cite{Pribiag:2007}. In addition
to studies in which the equilibrium orientation of the moments is in
the plane, devices have recently been produced in which the magnetic
anisotropy of one or more of the magnetic layers is manipulated so
that the magnetization points out of plane
\cite{Mangin:2006,Houssameddine:2007}.  Such devices may enable ways
to decrease the critical current for switching in memory devices, to
increase the switching speed \cite{Kent:2004}, or to produce improved
nano-oscillators.  Frequency- and time-domain measurements of the
different modes of spin-transfer-driven dynamics provide very fertile
ground for comparisons with sophisticated micromagnetic simulations of
the dynamics, as reported by Berkov and Miltat in their article.

An alternate measurement approach that has recently been exploited is
to apply a high frequency input current to drive resonant magnetic
precession and look at the static output voltage generated by mixing,
a technique referred to as spin-transfer ferromagnetic resonance
(ST-FMR)
\cite{Tulapurkar:2005,Sankey:2006,Sankey:2007,Kubota:2007,Kupferschmidt:2006,Kovalev:2007}.
These measurements, as applied to magnetic tunnel junctions, are
discussed in the article by Sun and Ralph.

{\em Effects of Non-Zero Temperature.}
Spin-torque devices have sufficiently small sizes that
temperature-induced fluctuations can have important effects on the
magnetic dynamics.  At room temperature, the reduced effective
damping that can be produced by spin transfer torque can enhance the
amplitude of thermal fluctuations even for currents well below the
$T=0$ critical current. These enhanced thermal
magnetic oscillations can be measured accurately in tunnel junctions
\cite{Petit:2007,Deac:2007}. In both tunnel junctions and metallic
devices, room temperature thermal fluctuations can also cause
magnetic switching to occur at currents that are well below the
intrinsic zero-temperature threshold current for spin-transfer-driven
instabilities (Eq.~(\ref{eq:critI})). In order to determine the
intrinsic critical current for switching,
it is necessary to make direct measurements of switching on the
typical time scale of ferromagnetic precession near 1~ns, or measure the
switching current as a function of temperature and extrapolate to
zero temperature, or measure the statistics of switching using current pulses
of various lengths and extrapolate to the ns scale.

The linewidths of
the microwave signals generated by spin-torque-driven magnetic
precession appear to be governed largely by thermal fluctuations that
produce deviations from perfectly periodic motion
\cite{Sankey:2005,Kim:2007}.  Point contact devices, in which magnetic
precession is excited in a local area of an extended magnetic film can
produce narrower linewidths than typical nanopillar devices
\cite{Rippard:2006}, perhaps because the micromagnetic exchange
coupling to the extended film makes them less susceptible to thermal
fluctuations. If the effects of thermal fluctuations can be reduced
or eliminated, it is likely that the ultimate limit on the linewidths
of spin-torque nano-oscillators will be chaotic dynamics of the
magnetization \cite{Lee:2004b,Berkov:2005,Yang:2007}.

In terms of theory, the role of temperature is an area
that is still under development.  Transport calculations and
determinations of the torque are typically done assuming zero
temperature.  This approximation is expected to be reasonable because
not much about the transport is expected to change with temperature
except for scattering rates. There have been a number of theoretical
studies of the temperature dependence based on the macrospin
approximation
\cite{Myers:2002,Li:2004,Apalkov:2005,Russek:2005,Xiao:2005}.
However, as the temperature increases, the macrospin approximation
becomes worse.  It is possible to include thermal effects into full
micromagnetic simulations, but the calculations become quite time
consuming and it is very difficult to capture meaningful statistics.

{\em On the Perpendicular Component of the Spin Torque Vector.}
In discussing the result of Eq.~(\ref{eq:intabsorb}),
{\it i.e}.\ that one expects essentially all of the
transverse spin angular momentum incident onto a
normal-metal/ferromagnet interface to be absorbed in producing the spin
torque, we made the point that this was only
approximately correct; it is not exact.  The most important caveat is
that classical averaging over the Fermi surface need not necessarily
eliminate all transport of transverse spin density away from the
interface, particularly for the reflected part of the scattering
wavefunction (or for very thin magnetic layers in transmission).  One
consequence is that the amplitude of the ``in-plane'' component of
the torque can differ somewhat
from Eq.~(\ref{eq:intabsorb}).  In addition, there is the possibility of
an additional contribution due to the spin torque in the form
\begin{eqnarray}
                 {\dot{\bf M}}_{\perp}
= \eta_{\perp}(\theta){\mu_{\rm B} I\over e \mathcal{V}}~
                 \hat{\bf M}\times\hat{\bf M}_{\rm fixed},
                 \label{eq:NsteffH}
\end{eqnarray}
oriented perpendicular to the plane defined by ${\bf M}$ and ${\bf
M}_{\rm fixed}$, rather than within this plane as in
Eq.~(\ref{eq:Nst}).
Note that here the symbol $\perp$ refers to the direction
perpendicular to the plane of the magnetizations as opposed to the
usage in Eq.~(\ref{eq:intabsorb}) where it refers to both components
of the spin current transverse to the free layer magnetization.
In Fig.~\ref{fig:precess}, this new component of torque would point in or
out of the page. The out-of-plane torque
component is sometimes referred to as an
``effective field'' contribution because
its form is similar to the torque that would result from a field
aligned with the fixed layer magnetization.  It can be viewed as a
consequence of a small amount of average precession into the ${\bf
\hat{y}}$ direction for reflected electrons in our toy models from
Section \ref{sec:stt}.

Calculations incorporating transmission and reflection coefficients
computed by {\it ab initio} techniques find that the perpendicular
component of the torque is small for the materials
generally used in metallic multilayer devices. For electrons of a
given energy, the out-of-plane component of spin transfer is predicted
to be less than 10~\%, and typically 1~\% to 3~\% of the in-plane component
\cite{Xia:2002,Stiles:2002a,Zwierzycki:2005}.  The final magnitude of
the bias-dependent part of the out-of-plane torque is expected to be smaller
still, due to a cancellation that arises when computing the bias dependence.
This cancellation is maximal in the case of a symmetric N/F/N/F/N
junction, where it can be understood from a simple symmetry argument.

\begin{figure}
              \centering
              \resizebox{0.6\columnwidth}{!}{%
              \includegraphics{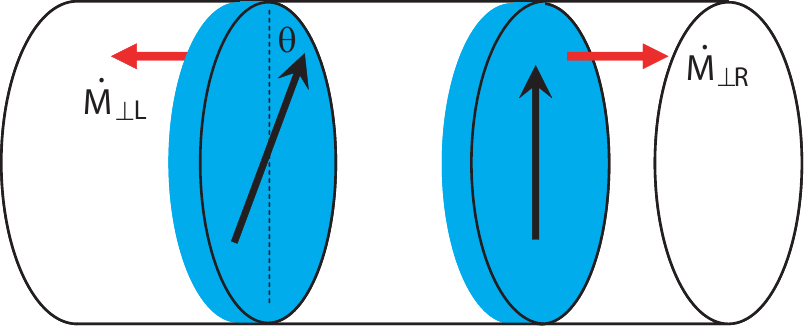}}
\caption{Sample geometry for the perfectly symmetric N/F/N/F/N device
       assumed in our analysis of the perpendicular component of the spin
       torque vector.  The perpendicular spin torques on the two magnetic
       layers are equal and opposite.
       }
              \label{fig:geomperp}
\end{figure}

Consider a junction whose layer structure is perfectly symmetric
about a plane at
the midpoint of the device, see Fig.~\ref{fig:geomperp}.  Assume that
the magnetic moments
of the two
ferromagnetic layers are oriented in the plane of the sample layers,
with an arbitrary angle $\theta$ between them. When a bias is applied to the
junction, the resulting spin
current density in each lead (each of the outer N layers) will be
aligned with the
magnetization of the neighboring ferromagnetic layer provided that
the electrons
that transmit into the lead from the rest of the devices are completely
classically dephased.  In this case, the spin
currents in the outer N layers are oriented
in the plane of the two magnetizations, and an out-of-plane spin
component of the
spin current density can exist only in the middle N layer.
If there are no sources or sinks of angular momentum except for the
spin torques
applied to the magnetic layers ({\it e.g.}, we assume that there is no angular
momentum lost to the excitation of short-wavelength spin wave modes),
then angular
momentum conservation allows the perpendicular component of spin torques on the
ferromagnetic layers to be determined solely from the out-of-plane spin current
traveling between the magnetic layers.  As a consequence, the
out-of-plane spin
torques on the
two magnetic layers must be equal and opposite. (To be clear, our
argument applies
only to the perpendicular component of the spin torque and not the in-plane
component, because the in-plane spin component of the spin-current
density is not
zero in the outer N layers, and hence the in-plane spin torques on the
ferromagnetic layers are not generally equal and opposite.)

For an applied bias such that electrons flow from left to right,
suppose that the
perpendicular spin torque on the right magnetic layer has the value
$N_{\perp}$.
Now imagine that the bias is reversed, so that electrons flow from
right to left.
Because of the symmetry
of the device, the spin current density in this case can
be determined from the first case simply by an appropriate rotation;
the answer is
that both
the direction of the out-of-plane spin component that flows between
the two layers and the direction of its flow are reversed, so that
(multiplying the two negative signs) the out-of-plane spin current
density is actually {\em the same} as in the first case.  Consequently, the
perpendicular spin torque acting on
the right magnetic layer due to electrons flowing right to left
will again be $N_{\perp}$, the same magnitude and
the same sign as for the perpendicular torque on the same (right)
magnetic layer
due electron flow from left to right.  It follows that the perpendicular spin
torque on a magnetic layer in a perfectly-symmetric junction
is an even function of
the bias $V$, as long as conservation of angular momentum can be applied to the
spin-transfer process and the classical dephasing processes are complete.
Consequently, there is
no contribution that is linear in $V$ at zero bias in a symmetric junction, in
spite of the fact that {\it ab initio} calculations show that
electrons at a given
energy incident onto a magnetic interface can give non-zero
contributions to the
perpendicular torque.

In a more microscopic picture, one can see how this result comes about by
considering a calculation in which the total perpendicular spin torque is
determined by integrating the contributions from all electrons
incident onto the
symmetric junction from both sides.  Following the logic of the
symmetry argument
above, the perpendicular torque due to incoming electrons in the energy range
$d\epsilon$ incident from the left should be the same, $N_{\perp}(\epsilon)
d\epsilon$, as for electrons of the same energy incident from the
right, as long
as we assume that the effect of the applied bias on the scattering potential is
sufficiently weak that the electron transmission and reflection amplitudes
do not depend explicitly on $V$.
For a symmetric junction, any applied voltage bias will offset the electron
chemical potential on the two sides of the device by $\pm
eV/2$.  Therefore, if we consider only elastic scattering processes, the
bias-dependent part of the perpendicular torque can be calculated by
subtracting the contributions of electrons in the energy range
$\epsilon_F - eV/2$
to $\epsilon_F$ that are no longer incident on the junction from one side, and
adding the contribution of the extra electrons in the range $\epsilon_F$ to
$\epsilon_F + eV/2$ that are incident from the other side:
\begin{eqnarray}
                &&{\dot{\bf M}}_{\perp}(V)- {\dot{\bf M}}_{\perp}(0) \nonumber\\
                &\propto&  - \int_{\epsilon{_F} -
eV/2}^{\epsilon_F} N_{\perp} (\epsilon) d\epsilon
                  +  \int_{\epsilon_F }^{\epsilon_F + eV/2 } N_{\perp} 
(\epsilon)
d\epsilon \nonumber\\
                &\approx& {d N_{\perp}(\epsilon_F) \over d\epsilon} \left(
{eV \over 2}
\right)^2.
\label{eq:Nstperp0}
\end{eqnarray}
The contributions from the states in the energy ranges $\epsilon_F - eV/2$ to
$\epsilon_F$ and $\epsilon_F$ to $\epsilon_F + eV/2$ cancel to first order in
$eV$, so that the lowest-order contribution to
the perpendicular spin torque should go only as $(eV)^2$, symmetric
in $V$ with no
linear term.  Even at non-zero biases the perpendicular component of the spin
torque is probably very small in most metallic multilayers. Predictions for
non-zero values of the perpendicular torque typically arise
only under the assumption
of coherent transport between ideal interfaces, but in real devices
this coherence
is generally not present
due to interface disorder
\cite{Waintal:2000,Zwierzycki:2005}.

For tunnel junctions, the bias-dependent part of the
out-of-plane spin torque is predicted to be
larger than for metallic multilayers \cite{Xia:2002,Theodonis:2006}.
It remains an open question as to how well our symmetry argument
applies to tunnel junctions, because the current may be dominated by a
small part of the Fermi surface.  In such systems, classical dephasing may not
be as complete and therefore in principle a small
linear-in-$V$ out-of-plane torque may remain
even for a
symmetric junction.
In addition, for tunnel junctions at high bias,
angular momentum loss from hot electrons to the excitation of
short-wavelength spin waves
might become significant, which would invalidate our arguments
because they are based on angular momentum conservation between
the conduction electrons and the approximately uniform magnetization mode.
Nevertheless, a tight-binding
calculation designed to model a symmetric magnetic tunnel junction
(in the absence of any short wavelength spin-wave excitations)
did predict a bias dependence
${\dot{\bf M}}_{\perp}(V)- {\dot{\bf M}}_{\perp}(0) \propto V^2$
\cite{Theodonis:2006}, consistent with the symmetry argument.
For a tunnel junction device with a
non-symmetric layer structure none of the symmetry arguments apply, and in this
case one should expect that there may be a perpendicular spin torque with a
linear dependence on $V$ near zero bias. The potential absence of
complete classical dephasing in magnetic tunnel junctions, the
consequences of spin-wave excitation by hot electrons, and the
effects of non-symmetric layer structures are all interesting
questions for future theoretical work, as is the effect of disorder
on the out-of-plane torque.

The experimental literature regarding the perpendicular component of
spin torque contains some contradictory results. Recent spin-transfer
ferromagnetic resonance
measurements (ST-FMR) on metallic NiFe/Cu/NiFe devices find no sign
of a perpendicular
torque at the level of 1~\% of the in-plane torque
\cite{Sankey:2007}, in agreement with the theoretical
expectations. However,
Zimmler {\it et al}.\ interpreted a measurement on Co/Cu/Co devices
\cite{Zimmler:2004} of a non-zero slope {\it vs}.\ current for the critical
magnetic field for switching $H_c(I)$ at 4.2 K as demonstrating a
perpendicular torque linear in $V$ with a magnitude about 20 \% of
the in-plane torque.
We are skeptical of this
interpretation, and suspect that the observation might be an artifact
of heating. The measured slope of $H_c(I)$ was not actually
constant as a function of $I$ as would be expected from a
perpendicular torque, but rather the slope of $H_c(I)$ was approximately zero
except for combinations of current polarity and magnetic orientation
for which the spin torque decreases the effective magnetic damping.  This
asymmetric rounding of the critical field line can be viewed as a
signature of thermally induced fluctuations
\cite{Myers:2002}, in that a
decreased effective damping can increase the amplitude of thermal
fluctuations and therefore decrease the measured switching field at
non-zero temperature.  Zimmler {\it et al}.\ attempted to correct for this
temperature effect in their analysis, but did not consider that the
temperature might be varying as a function of current due to heating
at 4.2 K.

Spin-transfer FMR measurements on symmetric magnetic tunnel junctions
differ from experiments on magnetic multilayers in that the
perpendicular torques found in the tunnel junctions are significant
\cite{Sankey:2007,Kubota:2007}.  In agreement with our symmetry
argument and with more detailed calculations
\cite{Theodonis:2006}, these experiments find that to good accuracy
${\dot{\bf M}}_{\perp}(V)- {\dot{\bf M}}_{\perp}(0) \propto V^2$ at
low bias for
symmetric tunnel junctions, with no contribution to the perpendicular
torque that is linear in $V$. An earlier report
by Tulapurkar {\it et al}.\ \cite{Tulapurkar:2005} of a
perpendicular torque linear in bias near
$V=0$ is now believed to be incorrect; the same group
has said more recently that this observation may
have been an artifact due to a spatially non-uniform magnetic state
\cite{Kubota:2007}.
The article by Sun and Ralph describes in more
detail the similarities and differences between spin torques in metal
multilayers and tunnel junctions.

{\em Comparison Between Theory and Experiments.}
While there is a general consensus about the correct approach for
calculating spin transfer torques in magnetic multilayers, there is as
of yet no fully quantitative comparison between theory and experiment
in metallic multilayers.
The primary difficulty is that the torques themselves are generally
not measured directly, but only the resulting dynamics as inferred
from the time dependent resistance.  Spin-transfer-driven
ferromagnetic resonance may eventually enable more direct quantitative
measurements, but so far this has not been achieved for the in-plane
component of the spin-torque in metal
multilayers. As described in the article by Berkov and Miltat,
attempting to infer the
dynamics of a magnetic sample from its time dependent
resistance is highly
non-trivial and may hide the details of the torque.  The problem is
particularly complicated because it is difficult to determine simply
from resistance measurements the degree to which spin-torque induced
excitations are spatially non-uniform.

The most sophisticated
calculations can make very good qualitative predictions concerning,
for example, the different types of dynamical modes (static or
precessional) that are observed experimentally.  However, they currently do
not match experiment well in more quantitative comparisons, for
instance in comparing to the current-dependent amplitude of the
microwave signals emitted in
the steady-state precession regime (see the article by
Berkov and Miltat). Another complication is that there is still some variation
experimentally between the behaviors of nominally identical samples.
This degree of variability has improved since spin transfer effects
were first measured, so that quantities like switching currents and
precession frequencies now show good reproducibility.  However, other
quantities, {\it e.g.}\ microwave linewidths and the positions of
transitions between different precession modes, are much more
variable.  The variability that still exists highlights the importance
of achieving improved control over materials and lithography.

\section{Domain Walls in Nanowires}\label{sec:wires}

In this section we provide a very brief introduction to current-induced
domain wall motion.  This section is brief compared to the
previous section because the articles by Beach, Tsoi, and Erskine, by
Tserkovnyak, Brataas, and Bauer, and by Ohno and Dietl together
provide a rather comprehensive introduction, review, and summary of
open questions for this topic.

Experimentally, current-induced domain wall motion is typically
studied in lithographically defined, narrow magnetic wires.  The wires
are frequently curved or the ends are designed to make it possible to
controllably introduce a domain wall into the wire using an applied
field.  In early experiments, the positions of the domain walls were
imaged using techniques such as Magnetic Force Microscopy (MFM)
\cite{Gan:2000,Tsoi:2003} or Scanning Electron Microscopy with
Polarization Analysis (SEMPA) \cite{Klaui:2003}.  Measuring the
positions before and after a current pulses allows estimates of the
wall velocities.  Alternatively, the locations of the wall could be
determined in real time using the Magneto-Optic Kerr Effect (MOKE)
\cite{Beach:2006} or electrically using GMR sandwich structures
\cite{Ono:2001,Grollier:2002} or through the extra resistance due to
anisotropic magnetoresistance (AMR) in a domain wall
\cite{Hayashi:2006}.  Using these various techniques, experimentalists
can determine the wall velocity as a function of current and
applied magnetic field and compare with theoretical predictions.  The
level of the comparison is discussed in detail in the article by
Beach, Tsoi, and Erskine.

To a first approximation, understanding current-induced domain wall
motion is quite simple \cite{Berger:1978}.  Assuming that the electron
spins adiabatically follow the magnetization direction, the divergence
of the spin current in Eq.~(\ref{eq:qdef}) gives a torque (actually a
torque density, but for the rest of the article we refer to it simply
as a torque) on the magnetization of $-({\bf v}_0 \cdot \bm\nabla){\bf
M}({\bf r})$.  Here, ${\bf v}_0$ is a vector in the current
direction with magnitude $v_0={P|j|\mu_B/ eM}$, where $P$ is the
polarization of the current. If the current is uniform, this torque
density simply translates the domain wall in the direction of electron
flow with a speed $v_0$.  There are several factors that complicate
this simple description.  These factors are described in the three
articles by Beach, Tsoi, and Erskine, by Tserkovnyak, Brataas, and
Bauer, and by Ohno and Dietl.

The degree to which the spins adiabatically follow the magnetization
has been computed in several models
\cite{Tatara:2004,Waintal:2004,Xiao:2006}.  The deviations are small
except for rather narrow domain walls.  When they are non-negligible,
there is an additional torque in the
${\bf M}\times({\bf v}_0 \cdot \bm\nabla){\bf M}$
direction.  This torque is referred to as a non-adiabatic torque
because it derives from the inability of the electron spins to
adiabatically follow the magnetization direction.

Much of the debate on the theoretical description of current-induced
domain wall motion is associated with how to describe the damping and
whether there is an additional torque in the direction of the
non-adiabatic torque \cite{Tatara:2004,Zhang:2004} that arises from
the same processes that lead to damping.  This torque, while not
related to a true non-adiabatic torque is still sometimes
called a non-adiabatic torque.  Alternatively, sometimes it is called the beta
term,
because $\beta$ is the dimensionless parameter that is frequently
used to characterize its strength.
This contribution is
extensively discussed by Tserkovnyak, Brataas, and Bauer.  One
technical aspect, whether the adiabatic spin transfer torque can be
derived from an energy functional, is addressed in the article by
Haney, Duine, N\'u\~nez, and MacDonald.  One of the present authors
believes that their argument is incorrect, but disagreement is one of
the things that makes an issue an open question.

Just as the dynamics of the free magnetic layers in nanopillars are
complicated by the possibility of spatially non-uniform magnetization
patterns, so
are
the dynamics of domain wall motion complicated by non-trivial wall
structures.  When the domain patterns distort, the motion is no longer
simple.  Domain walls distort in response to non-adiabatic torques,
damping, and the presence of non-uniformities in the sample.  The
non-uniformities are frequently described as pinning centers.  They
have been studied extensively in the context of field driven domain
wall motion as discussed by both Beach, Tsoi, and Erskine and Ohno and
Dietl.  Much experimental effort is spent on trying to minimize and
understand the effects of these pinning centers.

\section{Outlook}\label{sec:outlook}

In this article, we have tried to highlight some of the open questions
in the study of spin transfer torques and provide background material
for the succeeding articles.  As Katine and Fullerton describe, spin
transfer torques should start having commercial impact in the very
near future, but there are still important scientific issues to work
through.  It seems likely that tunnel junctions will be an important
system for applications of spin transfer torques.  Sun and Ralph
discuss what is known about spin transfer torques in magnetic tunnel
junctions as well as what we still need to learn.

The dynamics that result from spin transfer torques is the topic of
the articles by Berkov and Miltat and by Silva and Rippard.  The
former focuses primarily on the nanopillar geometry with a free layer
that is patterned to have a finite extent.  The latter focuses on the
fabricated nanocontact geometry in which the free layer is part of an
extended thin film.  Both show that the agreement between theory and
experiment is still incomplete.

The developments in current-induced domain wall motion are more recent
than those in magnetic multilayers and tunnel junctions, so
applications are still more uncertain.  As Beach, Tsoi, and Erskine
describe, the experiments are rapidly becoming more sophisticated and
more meaningful results are appearing.
Tserkovnyak, Brataas, and Bauer describe the rapid
progress in the theory of spin transfer torques in these systems.  It
is likely that experimental progress will challenge these theories
and more work will be required.

Most research on spin transfer torques has focused on transition metal
ferromagnets, but there are other potentially interesting materials systems.
Ferromagnetic semiconductors are potentially exciting because they
have magnetizations much smaller than transition metal magnets, so
that their moments might be manipulated using spin transfer from much
smaller currents than transition metal magnets. While the feasibility
of a room-temperature dilute ferromagnetic semiconductor remains to be
established, there are fascinating scientific questions to be
understood about these materials and their interactions with
current. Ohno and Dietl describe the measurements of current-induced
domain wall motion in a dilute magnetic semiconductor and a theory
that can be used to characterize the electrical and magnetic
properties of these systems.  Haney, Duine, N\'u\~nez, and MacDonald
describe a different way of computing spin transfer torques and apply
it to novel systems.  In particular, they raise the possibility of
large spin transfer torques in antiferromagnets.

\section{Acknowledgements}
We thank Piet Brouwer, Bob Buhrman, Paul Haney and Christian Heiliger
for valuable conversations about this manuscript, and many other colleagues
(too many to list) who have helped to educate us about spin transfer torques.
We also thank Julie Borchers, Arne Brataas, Yongtao Cui, Zhipan Li,
Jabez McClelland, Jacques
Miltat, Josh Parks, Vlad Pribiag, Kiran Thadani, Hsin-Wei Tseng, and Yaroslav
Tserkovnyak for reading the manuscript and offering
suggestions for improvement,
and Sufei Shi for help with figures.
The spin-torque work of DCR is supported by the
Office of Naval Research, the NSF (DMR-0605742 and EEC-0646547), and
DARPA.

\section{Notes on Corrections}
This version of the tutorial is modified from the
originally-published version.  ({\it i}) We have corrected algebraic
errors in Eq.\ (13) and modified the discussion in the following
paragraphs to correct a mistaken claim that the spin current density
in the toy model can be discontinuous at the interface.  We thank
Claire Baraduc for pointing out the error. ({\it ii}) We have
corrected an error in the units for the y axes in Figures 10a and 10b.


\begin{thebibliography}{100}
\bibitem{vcore}
A vortex core is truly only a singularity when the magnetization is
required to stay within the plane.  In real vortex cores, the
magnetization rotates out of the plane and the magnetization pattern
is continuous everywhere.
\bibitem{smallsize}
For a thin-film spin transfer device with a circular cross section, a
fixed layer
structure, and current flowing perpendicularly to the plane, the threshold for
spin-torque-driven excitations is given by a particular current
density, so that
the critical current scales $\propto r^2$, where $r$ is the device radius.
Reorientations produced by the current-induced Oersted field will
generally occur
at a particular field magnitude $H_c \propto I/r$, so that the
threshold current
for field-generated reorientations scales $\propto r$, approximately.
Therefore a
sufficiently small device radius is needed if one desires that the
threshold for
spin-torque excitations should occur at a lower current than thresholds for
magnetic-field-driven reorientations.
\end{thebibliography}
\end{document}